\documentclass[12pt]{article}
\usepackage[noBBpl,sc,osf]{mathpazo} 
\usepackage[scaled]{helvet} 
\usepackage{zlmtt} 
\usepackage[T1]{fontenc}
\usepackage{textcomp}
\normalfont

\usepackage[top=2cm]{geometry}

\usepackage[utf8]{inputenc}
\usepackage{amsthm}
\usepackage{amsmath}
\usepackage{tikz}
\usepackage{mathtools}
\usepackage{amssymb}
\usepackage{amsfonts}
\usepackage{parskip}
\usepackage{mathrsfs}
\usepackage{float}
\usepackage{setspace}
\usepackage{caption}
\usepackage{subcaption}
\usepackage[round]{natbib}
\newtheorem{lemma}{Lemma}
\newtheorem{fact}{Fact}
\newtheorem*{lemma A}{Lemma A}
\newtheorem{corollary}{Corollary}
\newtheorem{theorem}{Theorem}
\newtheorem{remark}{Remark}
\newtheorem{assumption}{Assumption}

\newtheorem{proposition}{Proposition}
\newtheorem{example}{Example}
\allowdisplaybreaks

\usepackage{verbatim}
\usepackage{hyperref}
   \usepackage{xcolor}
\hypersetup{colorlinks=true,linkcolor=blue,urlcolor=blue,citecolor=purple} 
\renewcommand{\b}[1]{\mathbf{#1}}
\renewcommand{\c}[1]{\mathcal{#1}}
\newcommand{\hb}[1]{\hat{\mathbf{#1}}}

\newcommand{\bb}[1]{\bar{\mathbf{#1}}}
\newcommand{\tb}[1]{\tilde{\mathbf{#1}}}
\newcommand{\sm}[2][K]{\sum_{{#2}\in\mathcal{#1}}}
\newcommand{\lmax}{\lambda_{max}}
\newcommand{\lmin}{\lambda_{min}}

\bibpunct[: ]{(}{)}{;}{a}{}{,}

\begin{document}
\title{{\bf Multi-activity Influence and Intervention}\footnote{
We thank the editor, an advisory editor, two anonymous referees as well as Francis Bloch, Yann Bramoulle, Arthur Campbell, George Charlson,  Andrea Galeotti, Sanjeev Goyal,  Rongzhu Ke,  Sudipta Sarangi, Fanqi Shi, Satoru Takahashi, Yiqing Xing, Yves Zenou, and seminar participants at 2021 Young Academics Networks Conference (Cambridge-INET), Zhejiang University, Monash University, and NUS. Zhou acknowledges support from Tsinghua Strategy for Heightening Arts, Humanities and Social Sciences: “Plateaus \& Peaks” (No. 2022TSG08102). The usual disclaimers apply.
}}
\author{Ryan Kor\thanks{Department of Economics, National University of Singapore, Singapore. E-mail: {e0004083@u.nus.edu}}\and Junjie Zhou\thanks{Corresponding author. School of Economics and Management, Tsinghua University, China. E-mail: {zhoujj03001@gmail.com}}}
\date{\today}
\maketitle

\begin{abstract}
Using a general network model with multiple activities, we analyse a planner’s  welfare maximising interventions taking into account within-activity network spillovers and cross-activity interdependence.  We show that the direction of the optimal intervention, under sufficiently large budgets, critically depends on the spectral properties of two matrices: the first matrix depicts  the social connections among agents, while the second one quantifies the strategic interdependence among different activities. In particular, the first principal component of the interdependence matrix determines   budget resource allocation across different activities, while the first (last) principal component of the network matrix  shapes the resource allocation  across different agents when network effects are strategic complements (substitutes).  We explore some comparative statics analysis with respective to model primitives and discuss several applications and extensions.
  \strut
 
 
\end{abstract}

\newpage


\section{Introduction}

Network models have made major contributions to the understanding of equilibrium activity in a variety of markets, such as criminal effort \citep{bcz,bcz2}, public goods \citep{bk,allouch2015private, elliott2019network}, and research and development \citep{goyal2001RD,klz}. These models take into account the significant spillover effects present in these activities to guide government intervention, see, for instance,  the key player problem in \cite{bcz}, and  the optimal targeting  interventions in \cite{ggg}. Nevertheless, it is important to consider that an intervention affects not only the intended activity, but also creates spillovers into other markets. For example, \cite{gold} showed that there are important strategic interactions between drug consumption and criminal activity, and so the predicted effects of an intervention that reduces criminal activity will be imprecise if the effects on the drug market is not taken into consideration. Therefore, in this paper, we discuss the implementation of government interventions when the players simultaneously participate in multiple activities.

We adopt the multiple activity network model in \cite{cyz}, and extend to the situation where the interactions between activities are not homogeneous. That is, we allow for the strength and type of interaction between each pair of activities to differ. In doing so, we define a matrix of strategic interdependence, and find that the eigendecomposition of this interdependence matrix, in combination with the network matrix,  plays a significant role in determining the optimal intervention and welfare. Our analysis generalizes the principal component analysis in \cite{ggg} to a multiple activity setting. In particular, we obtain analogous ``simple'' interventions that the planner can adopt to obtain an asymptotically optimal welfare.

We also establish results on the allocation of the planner's budget when the budget is large, both across activities, and across agents within an activity. We show that the two allocations are in a sense independent of each other, and are determined solely by the eigenvectors of the two  matrices mentioned before - the matrix of strategic interdependence and the network adjacency matrix.  More precisely, we show that the first principal component of the interdependence matrix determines   budget resource allocation across different activities, while the first (last) principal component of the network matrix  shapes the resource allocation  across different agents when network effects are strategic complements (substitutes). We also perform some comparative statics analysis to obtain monotonicity results in the case where activities are complements.

To broaden the scope of our study, we also consider the loss in welfare due to the restriction in intervention (called partial intervention) whereby the planner is unable to intervene in certain activities. This could be due to a lack of infrastructure to facilitate effective targeted intervention. For example, using the example of drug consumption above, there may be underground supply chains that the planner is unable to control. Consequently, the planner could decide to focus on intervening in the dimension of criminal activity, and indirectly influence drug consumption through the strategic interdependence. In this paper, we show that when the activities are complements and the budget is large, a restriction in the intervention space will lead to a percentage loss in total social welfare among the players, and this welfare loss is proportional to the number of restricted activities. On the other hand, there is no loss in the case of substitutable activities as long as at least two activities are available for intervention. These results showcase the difference between complementary and substitute activities, where we see that increasing the intervention space is more important when the activities are complements. In contrast, when the budget is small, there are decreasing marginal benefits to the number of activities that the planner intervenes in regardless of the type of strategic interactions. This trend thus lies between the extreme cases observed when the planner's budget is large, and provides another consideration for policy implementation.

Finally, we find that the impact of a restriction in the planner's intervention is larger when the intensity of the strategic interactions or the network spillovers are strong. Intuitively, the maximum welfare returns to the budget is larger in these cases due to the greater strategic relationships in effort levels that the planner can exploit with a suitable choice of intervention, but the planner is unable to do so under a restricted intervention space. Similarly, the effect of a restriction is also greater for denser graphs when there are strategic complementarities among the agents due to the network spillovers. On the other hand, the effect of an increase in interconnectedness between the agents is ambiguous when there are strategic substitutabilities instead, as the additional linkages can result in a decrease in consumption.

We also consider several extensions to our model. Firstly, we relax our assumptions to allow for the network spillovers to be different in each activity. We find that our main results on the spectral properties of the optimal intervention are still largely applicable with some modifications. In particular, the first and last principal components of the adjacency matrix still feature prominently in the optimal intervention. We also studied the problem of nonnegative interventions, which is where the planner is only able to shift incentives in one direction. This is relevant in situations where a planner may find it impractical to decrease accessibility to an activity such as education, and thus is only able to incentivise further participation. While the problem is in general shown to be NP-hard, our simulations offered some intuition to the optimal interventions, and may still guide policy direction. Finally, we offer an alternative interpretation of our model in a monopoly setting. A price discriminating monopolist distorts the market in a manner equivalent to the targeted interventions we study, but aims to maximize its total profit instead of the consumers' welfare. We show that the results in \cite{cbo} and \cite{bloch2013pricing} can be generalized to our case of a multiple-good market.

Throughout our analysis, we assume that the underlying network structure is the same across all activities.\footnote{In one extension, \cite{cyz} consider a  model with multiple distinct networks.} While this seems restrictive, as a first exploration into interventions on multiplex networks, our model still lends itself to various economic applications. For example, under the same criminal social network, players may choose their involvement in both drug consumption and criminal activity. A planner could then choose to target each activity separately---the planner could decrease criminal activity by increasing law enforcement and police presence, or reduce drug consumption by an intervention in the drug market. Since the players enjoy complementary effects from participation in both activities \citep{gold}, our results suggest that the planner should allocation equal amounts into the intervention on both activities. A planner who ignores the cross-activity interactions and only intervenes in one dimension would incur a significant decrease in optimal welfare.

Another multiple activity setting, where consumers instead participate in substitutable activities, can be found in a social network with the consumers choosing their participation levels among numerous video-sharing and networking platforms such as Facebook, Instagram, and Tiktok. Our results thus imply that a planner who wants to maximize utility from these could simply focus their efforts on promoting the usage of just one or two applications.

\subsection*{Related Literature}
There is a range of other literature in both the multi-activity dimension, as well as the study of interventions. \cite{bcz} introduced the seminal single-activity network model, showing that the equilibrium activity level is related to the Katz-Bonacich centralities of the network.\footnote{See \cite{SET2022} for recent developments in network models with nonlinear responses.} \cite{bd} extends the single-activity model to the case of agents participating in two perfectly substitutable activities, and \cite{cyz} generalises the analysis to arbitrary strategic interactions between multiple activities. \cite{walsh} also considers a network model where agents invest in two public goods. \cite{goyal2017aer} develop a multi-activity model where individuals in a social network choose whether to participate in their network and whether to participate in the market.

In the area of network interventions, a wide variety of methods have been proposed in \cite{val} for a planner to conduct. Structural interventions, such as in \cite{bcz}, \cite{gl}, \cite{cs} and \cite{SZZ2021structural}, adjust activity levels through modifying the network structure. The creation or deletion of links in such interventions affects the centralities of the agents, which leads to a change in the equilibrium. Other papers such as \cite{dem} and \cite{ggg} have considered characteristic interventions, where the planner instead modifies the agents' intrinsic valuations of the activities.\footnote{\cite{kor2022joint} analyze joint interventions in  both characteristics of nodes and weights on the links that connect nodes.}
 Such characteristic interventions are the main focus of our paper.

A closely related field of research is that of discriminatory pricing within a network of consumers, where a decrease in price for a consumer has a similar effect as an increase in intrinsic utility. \cite{cyz2,O2p2022} study the optimal pricing for both monopolies and oligopolies, as well as the implications on total welfare, while \cite{uz} considers a version with product varieties. \cite{fg} allows for incomplete information of the network structure.

However, to the best of our knowledge, the previous literature has only analysed characteristic interventions in a single activity. Therefore, this is the first attempt in including interventions into a multiple activity network model. Our paper follows and extends the methodology in \cite{ggg} to analyse a novel set of issues that appear only in the setting with multiple activities. For instance, we study the effects of restrictions on the planner's intervention space, and highlight the importance of cross-activity interdependence in shaping the optimal interventions. Our paper and \cite{ggg}, together, provide a more complete theory about targeting interventions in networks with complex interactions across agents and across activities. 

The remainder of this paper is organized as follows. Section \ref{sect-2} introduces the model, as well as the key assumption and notations used in this paper. Section \ref{sect-3} solves for the equilibrium and analyses some comparative statics. Section \ref{sect-4} broadens the scope of the model by considering a case where the planner is limited in its interventions, and Section \ref{sect-5} explores several other extensions which apply the model in different contexts. Finally, Section \ref{sect-6} concludes the paper. Appendix \ref{sect-A} presents some preliminary mathematical results, and Appendix \ref{sect-B} provides the proofs of the results presented in this paper.

\section{Model}\label{sect-2}

We introduce a general multi-activity  network model and analyze the optimal interventions.

\underline{\bf Network}
Consider a network game where a set of agents \(\mathcal{N}=\{1,2,\cdots,n\}\) participate in a set of activities \(\mathcal{K}=\{1,2,\cdots,k\}\), with \(k\geq2\).\footnote{When \(k=1\), the network model reduces to the single activity model discussed in \cite{bcz} and \cite{bk}, with the optimal intervention characterized  thoroughly by \cite{ggg}.} Let \(\b G=(g_{ij})\) be the adjacency matrix of the network, allowing for arbitrarily weighted graphs, so that \(g_{ij}\in\mathbb R_+\) for all \(i,j\in\c N\).\footnote{For unweighted graphs, we have the specific case \(g_{ij}\in\{0,1\}\) for all \(i,j\in\c N\).} We also assume that \(g_{ii}=0\) for all \(i\in\mathcal{N}\), that is, there are no self-loops. Further assume that \(\b G\) represents an undirected network, so that \(\b G\) is symmetric, i.e., \(g_{ij}=g_{ji}\) for all \(i,j\in\mathcal{N}\).

\underline{\bf Payoffs}
Each agent \(i\) chooses actions \(\b x_i=(x_i^1, x_i^2,\cdots, x_i^k)\in\mathbb R^k\) simultaneously, where each \(x_i^s\) represents agent \(i\)'s level of participation in activity \(s\in\c K\).  We suppose that the payoff to each agent \(i\) is given by the utility function\footnote{This utility specification  extends that in  \cite{cyz}, which assumes $\beta_{ij}=\beta,\forall i\neq j$. } 
\begin{equation}U_i(\b x_i;\b x_{-i},\b a)= \underbrace{\sum_{s\in\mathcal{K}}a_i^sx_i^s-\left(\frac{1}{2}\sum_{s\in\mathcal{K}}(x_i^s)^2+\frac{1}{2}\sum_{s\in\mathcal{K}}\sum_{\substack{t\in\mathcal{K}\\t\neq s}}\beta_{st}x_i^sx_i^t\right)}_\text{private utility}+\underbrace{\delta\sum_{s\in\mathcal{K}}\sum_{j\in\mathcal{N}}g_{ij}x_i^sx_j^s}_\text{network spillovers}.\label{eq-1}\end{equation}

We explain the utility function \eqref{eq-1} term by term. For each \(i\in\mathcal{N}\) and \(s\in\mathcal{K}\), the parameter \(a_i^s\) represents player \(i\)'s intrinsic marginal utility from activity \(s\). The cost of actions consists of two parts: the first is the sum of the quadratic term \(\frac{1}{2}(x_i^s)^2\) over $s$; the second is the sum of interaction terms $\beta_{st} x_i^sx_i^t$  over different activities  $s\neq t$. Here \(\beta_{st}\) represents the degree of strategic substitutability or complementarity between the activities \(s\) and \(t\): \[\partial^2 U_i/\partial x_i^s\partial x_i^t=-\beta_{st},\] so a positive \(\beta_{st}\) corresponds to the case where the activities are substitutes, while a negative \(\beta_{st}\) corresponds to the case where the activities are complements. When \(\beta_{st}=0\), there are no direct interactions between the activities. Note that without loss of generality, we can let \(\beta_{st}=\beta_{ts}\), otherwise we can replace them with their average without changing the utility function. We will impose some regularity condition on the $\beta_{st}$ to guarantee convexity of the utility function.

Lastly, the third term captures the total network externalities enjoyed by agent \(i\), where \(\delta\) represents the strength of the network externalities. We assume that the strength of these externalities is the same for each activity. Notice that, for each $s$, \[\partial^2U_i/\partial x_i^s\partial x_j^s=\delta g_{ij},\] so the sign of \(\delta\) determines the strategic interaction between agents, and an increase in \(|\delta|\) reflects an increase in the intensity of these network spillovers. \cite{bcz} investigates the case of \(\delta>0\), representing strategic complementarities between the agents in a model of a network with a single activity. On the other hand, \(\delta<0\) corresponds to strategic substitutability between agents, as in the model of a public good network by \cite{bk}. When \(\delta=0\), there are no network effects and each agent receives utility only based on their own choices of actions. Both \cite{bcz} and \cite{bk} consider a single activity network model. Note that in our setting, the network effects aggregate over different activities.

Throughout this paper, we reserve the indices \(i,j\) to represent players and \(s,t\) to represent activities.
For convenience, we introduce the following notation:\[\b a^s=\begin{bmatrix}a_1^s\\\vdots\\a_n^s\end{bmatrix}\in\mathbb R^n,\ \b a=\begin{bmatrix}\b a^1\\\vdots\\\b a^k\end{bmatrix}\in\mathbb R^{kn},\ \b x^s=\begin{bmatrix}x_1^s\\\vdots\\x_n^s\end{bmatrix}\in\mathbb R^n,\ \text{ and }\b x=\begin{bmatrix}\b x^1\\\vdots\\\b x^k\end{bmatrix}\in\mathbb R^{kn}.\]

\underline{\bf Targeted Intervention}
We let \(\hb a\) be the original vector of the agents' marginal utilities. The social planner intervenes by shifting \(\hb a\) to a new vector \(\b a\), with the aim of maximizing the total social welfare \[W(\b a)=\sum_{i\in\mathcal{N}}U_i(\b x_i^*(\b a);\b x_{-i}^*(\b a),\b a),\] where \(\b x^*(\b a)\) denotes the equilibrium activity level given the planner's choice of \(\b a\) (see Proposition \ref{pr-1} for the equilibrium characterization). In other words, \(\b x_i^*(\b a)\) is a best response to \(\b x_{-i}^*(\b a)\) for all \(i\). This intervention comes at a cost to the planner, which we will model using a quadratic cost as \(\|\b a-\hb a\|^2\) following \cite{ggg}. We assume that the planner can incur a maximum expenditure of \(C\in\mathbb R^+\), and thus solves the constrained optimization problem 
\begin{align*}
    \max_{\b a\in\mathbb R^{kn}} \quad &\sm[N]iU_i(\b x_i^*(\b a);\b x_{-i}^*(\b a),\b a)\stepcounter{equation}\tag{\theequation}\label{eq-objective}\\
    \text{s.t.} \quad 
    &(\b a-\hb a)^T(\b a-\hb a)\leq C,&\text{(Budget constraint)}\\
    &\b x^*_i(\b a)\in\underset{\b x_i\in\mathbb R^k}{\text{argmax}}\ U_i(\b x_i;\b x_{-i}^*(\b a),\b a)\text{ for all }i\in\c N.&\text{(Agents' equilibrium)}
\end{align*}
We write the optimal choice of intervention as \(\b a^*(C)\), with a corresponding total welfare of \(W(\b a^*(C))=W^*(C)\).

\subsection{Assumptions and Notation}
We first define some standard matrices. Let \(\b I_p\) be the \(p\times p\) identity matrix, \(\b 0_p\) be the \(p\times p\) matrix of zeroes, and \(\b J_p\) be the \(p\times p\) matrix of ones.

Given a real symmetric matrix \(\b M\), define \(\lmax(\b M)\) and \(\lmin(\b M)\) be the largest and smallest eigenvalues of \(\b M\) respectively. Also denote their respective eigenspaces by \(E_{max}(\b M)\) and \(E_{min}(\b M)\). Note that since \(\b G\) is nonnegative, by the Perron-Frobenius theorem, we have that \(\lmax(\b G)\geq0\geq\lmin(\b G)\geq-\lmax(\b G)\).

Define the strategic interdependence matrix 
\[\b \Phi=\begin{pmatrix}0&\beta_{12}&\cdots&\beta_{1k}\\\beta_{21}&0&\cdots&\beta_{2k}\\\vdots&\vdots&\ddots&\vdots\\\beta_{k1}&\beta_{k2}&\cdots&0\end{pmatrix},\]
and further write $$ \tb \Phi=\b I_k+\b \Phi.$$
 We observe that \(\b\Phi\) and \(\tb\Phi\) are both symmetric, \(\lmin(\tb \Phi)=1+\lmin(\b \Phi)\), and \(E_{min}(\tb \Phi)=E_{min}(\b \Phi)\). 

Throughout the paper, we impose the following assumption:
\begin{assumption}
\label{ass-1}
 \(1+\lmin(\b \Phi)-\lmax(\delta\b G)>0\).
\end{assumption}

This assumption is equivalent to  $\lmin(\tb \Phi)-\lmax(\delta\b G)>0$, which specifies a sufficient condition to ensure that the underlying network game among agents has a unique Nash equilibrium for any $\b a$.
 Note that Assumption \ref{ass-1} directly implies that \(\tb \Phi\) is positive definite. When the network externalities are positive, \(\delta>0\) and \(\lmax(\delta\b G)=\delta\lmax(\b G)\geq 0\). On the other hand, when the network externalities are negative, \(\delta<0\) and \(\lmax(\delta\b G)=\delta\lmin(\b G)\geq 0\). In all cases, we have \(\lmin(\tb \Phi)>\lmax(\delta\b G)\geq 0\), so \(\tb \Phi\) is positive definite. Thus each agent's payoff is concave in her own strategy.
 Assumption \ref{ass-1} makes sure the network effects are not too strong.  

Assumption \ref{ass-1} generalizes the condition on the spectral radius of \(\b G\) as found in previous literature. Suppose the interaction between any pair of activities is the same, so that \(\beta_{st}=\beta\) for all \(s,t\). Then the distinct eigenvalues of \(\b \Phi\) are \((k-1)\beta\) and \(-\beta\). Therefore, Assumption \ref{ass-1} reduces to the requirement that \[1-\beta-\lmax(\delta\b G)>0\text{ and }1+(k-1)\beta-\lmax(\delta\b G)>0,\]
which is equivalent to the condition stated in \cite{cyz}. As a further specialization, when the activities are independent and \(\beta_{st}=0\) for all \(s,t\), we have \(\b \Phi=\b 0_k\) and \(\lmin(\b \Phi)=0\). Thus Assumption \ref{ass-1} reduces to \(1-\lmax(\delta\b G)>0\), which is further simplified to $\delta <\frac{1}{\lmax(\b G)}$ when $\delta>0$ (see \cite{bcz}), and $|\delta| <-\frac{1}{\lmin(\b G)}$ when $\delta<0$  (see, for instance,  \cite{bk}).

Finally, for any matrices \(\b A_{m\times n}\) and \(\b B_{p\times q}\), let \(\otimes\) denote the Kronecker product, where \(\b A\otimes\b B\) represents the block matrix \[\begin{bmatrix}a_{11}\b B&\cdots&a_{1n}\b B\\\vdots&\ddots&\vdots\\a_{m1}\b B&\cdots&a_{mn}\b B\end{bmatrix}.\]

\section{Analysis}\label{sect-3}
\subsection{Optimal Intervention}

We first derive the equilibrium activity profile among agents and the aggregate equilibrium payoff, fixing  $\b a$.
\begin{proposition}\label{pr-1}
Suppose Assumption \ref{ass-1} holds.  There exists a unique equilibrium in which
the agents choose activity levels
 \[\b x^*(\b a)=[\tb \Phi\otimes\b I_n-\b I_k\otimes\delta\b G]^{-1}\b a,\] 
 for a total  equilibrium welfare of \[W(\b a)=\frac{1}{2}\b a^T[\tb \Phi\otimes\b I_n-\b I_k\otimes\delta\b G]^{-1}(\tb \Phi\otimes\b I_n)[\tb \Phi\otimes\b I_n-\b I_k\otimes\delta\b G]^{-1}\b a.\]
\end{proposition}
For simplicity, we will write 
\begin{equation}
\b P=\frac{1}{2}[\tb \Phi\otimes\b I_n-\b I_k\otimes\delta\b G]^{-1}(\tb \Phi\otimes\b I_n)[\tb \Phi\otimes\b I_n-\b I_k\otimes\delta\b G]^{-1},
\label{eq-P}
\end{equation}
so that \(W(\b a)=\b a^T\b P\b a\).
Proposition \ref{pr-1} generalises the equilibrium results found in previous literature to the case of multiple heterogeneous activities, and the proof can be found in Appendix \ref{sect-B}. Relating our results to the existing literature, we illustrate Proposition \ref{pr-1} with a few simple cases.

\underline{\bf One activity} When there is only one activity, then \(k=1, \tb \Phi=\b I_1\), so \[\b x^*(\b a)= (\b I_n-\delta\b G)^{-1}\b a,\mbox{ and } W(\b a)=\frac{1}{2}\b a^T(\b I_n-\delta\b G)^{-2}\b a.\]
This reduces to the equilibrium obtained in \cite{bcz}, where activity levels are equal to the Katz-Bonacich centralities of each agent, and welfare proportional to the squared activity levels. \cite{ggg} studies the optimal targeted intervention under this framework.

\underline{\bf Two activities} When there are two activities ($k=2$), we have \(\tb\Phi=\begin{pmatrix}1&\beta\\\beta&1\end{pmatrix}\), where we let \(\beta_{12}=\beta_{21}=\beta\) by the symmetry assumption. We can expand the tensor products to obtain the equilibrium activity levels\[\b x^*(\b a)=\left(\begin{bmatrix}\b I_n&\beta\b I_n\\\beta\b I_n&\b I_n\end{bmatrix}-\begin{bmatrix}\delta\b G&\b 0\\\b 0&\delta\b G\end{bmatrix}\right)^{-1}\b a=\begin{bmatrix}\b I_n-\delta\b G&\beta\b I_n\\\beta\b I_n&\b I_n-\delta\b G\end{bmatrix}^{-1}\b a.\]
Writing \(\b M^+=[(1+\beta)\b I_n-\delta\b G]^{-1}\) and \(\b M^-=[(1-\beta)\b I_n-\delta\b G]^{-1}\), the solution (see \cite{cyz}) is given by 
\[\begin{bmatrix}\b x^1\\\b x^2\end{bmatrix}=\frac{1}{2}\begin{bmatrix}\b M^+(\b a^1+\b a^2)+\b M^-(\b a^1-\b a^2)\\\b M^+(\b a^1+\b a^2)-\b M^-(\b a^1-\b a^2)\end{bmatrix}.\]
Therefore, \(\b M^+\) determines the total action over both activities, while \(\b M^-\) determines the difference in action between both activities. Finally, the total welfare is \begin{align*}
W(\b a)&=\frac{1}{2}\b a^T\begin{bmatrix}\b I_n-\delta\b G&\beta\b I_n\\\beta\b I_n&\b I_n-\delta\b G\end{bmatrix}^{-1}\begin{bmatrix}\b I_n&\beta\b I_n\\\beta\b I_n&\b I_n\end{bmatrix}\begin{bmatrix}\b I_n-\delta\b G&\beta\b I_n\\\beta\b I_n&\b I_n-\delta\b G\end{bmatrix}^{-1}\b a\\&
=\frac{1}{2}\b x^T\begin{bmatrix}\b I_n&\beta\b I_n\\\beta\b I_n&\b I_n\end{bmatrix}\b x\\&=\frac{1}{2}[(\b x^1)^T\b x^1+2\beta(\b x^1)^T\b x^2+(\b x^2)^T\b x^2]\\&=\frac{1}{4}[(1+\beta)(\b a^1+\b a^2)^T(\b M^+)^2(\b a^1+\b a^2)+(1-\beta)(\b a^1-\b a^2)^T(\b M^-)^2(\b a^1-\b a^2)].\end{align*}
The presence of the cross-activity interaction term \(2\beta(\b x^1)^T\b x^2\) means that the total welfare is no longer equal to the squared activity levels. Instead, welfare can again be decomposed into two parts involving the total marginal utilities and the difference in marginal utilities, in a manner similar to the equilibrium \(\b x^*\).

However, further equilibrium analysis via expanding the relevant tensor products will not be feasible for large number of activities. Instead, we reformulate the planner's optimization problem \eqref{eq-objective} using the above equilibrium characterization to obtain a constrained quadratic maximization problem:
\begin{align}
    \max_{\b a\in\mathbb R^{kn}} \quad &\b a^T\b P\b a\label{eq-qp}\\
    \text{s.t.} \quad &(\b a-\hb a)^T(\b a-\hb a)\leq C.\notag
\end{align}

We then exploit general results of constrained quadratic programming, which we collectively state in Lemma \ref{lem-1}.
Lemma \ref{lem-1} is applicable both in  our paper and  \cite{ggg}, given the structural similarity in the underlying programs.  

\begin{lemma}\label{lem-1}
Let \(\b S\) be a positive definite matrix, and \(\b v\) a vector in \(\mathbb R^n\). Then the solution \(\b x^*\) to the maximization problem 
\begin{align*}
    \max_{\mathbf x\in\mathbb R^{n}} \quad &V(\b x)=\mathbf x^T\mathbf{Sx}+\b v^T\b x\\
    \text{s.t.} \quad 
    &\b x^T\b x\leq C
\end{align*}
satisfies\\
(a) \(\lim_{C\to\infty}\frac{V(\b x^*)}{C}=\lmax(\b S)\), and for any unit vector \(\b u\in E_{max}(\b S)\), \(\lim_{C\to\infty}\frac{V(\sqrt C\b u)}{V(\b x^*)}=1\), \\
(b) \(\lim_{C\to\infty}\frac{\|\mathrm{proj}_{E_{max}(\b S)}\b x^*\|}{\|\b x^*\|}=1\).\\
Furthermore, if \(\b v\neq\b 0\), we have\\
(c) \(\lim_{C\to0}\frac{V(x^*)}{\sqrt C}=\|\b v^T\|\),\\
(d) \(\lim_{C\to0}\frac{\|\text{proj}_{\b v}\b x\|}{\|\b x\|}=1\).
\end{lemma}

Lemma \ref{lem-1} shows that as \(C\to\infty\), the solution to \eqref{eq-qp} simply hinges on the largest eigenvalue of \(\b P\) and its corresponding eigenspace. Consequently, our problem reduces to obtaining the spectral decomposition of $\b P$, which can be rewritten as \[\b P=\frac{1}{2}[\tb \Phi^{\frac{1}{2}}\otimes\b I_n-\tb \Phi^{-\frac{1}{2}}\otimes\delta\b G]^{-2}.\footnote{Given a positive definite matrix \(\b M\), the square root of \(\b M\), written \(\b M^{\frac{1}{2}}\), is the unique positive definite matrix satisfying \((\b M^{\frac{1}{2}})^2=\b M\).}\] Here we make use of the fact that \(\tb \Phi^{\frac{1}{2}}\otimes\b I_n\) and \(\tb \Phi^{-\frac{1}{2}}\otimes\delta\b G\) are simultaneously diagonalizable since they commute. Let \(\b Q\b \Lambda\b Q^{-1}\), \(\b R\b \Sigma\b R^{-1}\) be spectral decompositions of \(\tb\Phi\) and \(\delta\b G\), respectively. Then \begin{align*}\b P&=\frac{1}{2}[(\b Q\otimes\b R)(\b \Lambda^{\frac{1}{2}}\otimes\b I_n-\b \Lambda^{-\frac{1}{2}}\otimes\b \Sigma)(\b Q\otimes\b R)^{-1}]^{-2}\\&=\frac{1}{2}(\b Q\otimes\b R)^2(\b \Lambda^{\frac{1}{2}}\otimes\b I_n-\b \Lambda^{-\frac{1}{2}}\otimes\b \Sigma)^{-2}(\b Q\otimes\b R)^{-2},\end{align*}
so the eigenvalues of \(\b P\) are the entries of the diagonal matrix \[\frac{1}{2}(\b \Lambda^{\frac{1}{2}}\otimes\b I_n-\b \Lambda^{-\frac{1}{2}}\otimes\b \Sigma)^{-2},\]
which are \begin{equation}\frac{1}{2}[\lambda_i(\tb\Phi)^{\frac{1}{2}}-\lambda_i(\tb\Phi)^{-\frac{1}{2}}\lambda_j(\delta\b Q)]^{-2}\text{ for }i=1\cdots k\text{ and } j=1\cdots n.\label{eq-eigenvalues}\end{equation}

Equation \eqref{eq-eigenvalues} fully characterizes the spectral properties of \(\b P\). 

Employing Lemma \ref{lem-1}, we now state our first main result regarding the planner's optimal intervention policy. 
\begin{theorem} \label{th-1}
Suppose Assumption \ref{ass-1} holds.\footnote{The expression in Theorem \ref{th-1}(a) is well defined when Assumption \ref{ass-1} holds. When \(\lmin(\tb\Phi)-\lmax(\delta\b G)\leq0\), the network effects are too strong, so the optimal welfare is unbounded and no Nash equilibrium exists.}\\
(a) As the planner's budget \(C\to\infty\), \[\frac{W^*(C)}{C}\to\frac{\lmin(\tb \Phi)}{2(\lmin(\tb \Phi)-\lmax(\delta\b G))^2}\equiv \Gamma(\b \Phi,\delta\b G).\]
(b) Furthermore, if \(\b u\in E_{min}(\b \Phi),\b v\in E_{max}(\delta\b G)\) are unit vectors, then the intervention \(\tb a(C)=\sqrt C\b u\otimes\b v+\hb a\) is asymptotically optimal in the sense that \[\lim_{C\to\infty}\frac{W(\tb a(C))}{W^*(C)}=1.\]
\end{theorem}
Theorem \ref{th-1} characterizes the  growth rate of  the optimal welfare $W^*$ and provides a \emph{simple} intervention $\tb a$ to achieve $W^*$ asymptotically under large budgets.
Our result implies that the planner does not need to identify the prior marginal utilities to be able to implement an asymptotically efficient intervention. 
The term \(\Gamma(\b \Phi,\delta\b G)\) measures the marginal return of the planner's budget, as
 \[
\lim_{C\to\infty} W^{*'}(C)=\lim_{C\to\infty}\frac{W^*(C)}{C}=\Gamma(\b \Phi,\delta\b G)
\]
by Theorem \ref{th-1} and L'Hospital's rule.

Theorem \ref{th-1} extends and generalises the results obtained in \cite{ggg} with a single activity to a network setting with multiple activities, and where the interaction between each pair of activities can be arbitrary. Here, in addition to the adjacency matrix  $\b G$ of the network, there is another matrix \(\b \Phi\), describing the strategic interactions between the multiple activities, that is crucial in determining the optimal intervention and welfare. 

\begin{proposition}\label{pr-2}
Let \(\Gamma(\b \Phi,\delta\b G)\) be defined as in Theorem \ref{th-1}. Then
\[\frac{\partial \Gamma(\b \Phi,\delta\b G)}{\partial\lmax(\delta\b G)}>0 \text{ and } \frac{\partial \Gamma(\b \Phi,\delta\b G)}{\partial\lmin(\b \Phi)}<0.\]
\end{proposition}

Note that  $\Gamma(\b \Phi,\delta\b G)$ depends only on the smallest eigenvalue of $\lmin(\tb \Phi)$ and the largest eigenvalue of $\lmax(\delta\b G)$. From Theorem \ref{th-1}(a), \(\Gamma(\mathbf\Phi,\delta\b G)\) is equal to the asymptotic return to the budget, \(W^*(C)/C\). Therefore, a larger value of \(\Gamma(\mathbf\Phi,\delta\b G)\) would imply a greater effect of the budget on welfare gain. Consequently, Proposition \ref{pr-2} shows that when the budget is sufficiently big, the planner's intervention will be more effective in the following two cases:
\begin{enumerate}
\item[(i)]
 \(\delta\b G\) has a large spectral radius, so the maximum social multiplier effect that can be caused by the network externalities is large, which the planner exploits by choosing the intervention in the direction of the first principal component.
 \item
[(ii)] \(\mathbf \Phi\) has a small last eigenvalue, so the minimum cost incurred by the cross-activity terms of the utility function is small and the planner can conduct a more cost-efficient intervention by intervening in the direction of the last principal component.
\end{enumerate}

For ease of notation, we introduce another assumption:
\begin{assumption} \label{ass-2}
 \(E_{min}(\b \Phi)=\text{span}(\b u)\) and \(E_{max}(\delta\b G)=\text{span}(\b v)\) for some unit vectors \(\|\b u\|=\|\b v\|=1\).
 \end{assumption}

Assumption \ref{ass-2} implies that the dimensions of \(E_{min}(\b \Phi)\) and \(E_{max}(\delta\b G)\) are both one, and holds generically. This ensures that the optimal intervention is essentially unique,\footnote{There can be another optimal intervention obtained by multiplying by \(-1\), but this transformation does not affect our subsequent results.} and \(\lim_{C\to\infty}\rho(\b u\otimes\b v,\b a^*)=\pm 1\), where \(\rho\) represents the cosine similarity\footnote{See \cite{ggg}, or Lemma \ref{lem-1}(b) in Appendix \ref{sect-A}.} \[\rho(\b x,\b y)=\frac{\b x^T\b y}{\|\b x\|\|\b y\|}.\]

We define the components of the planner's expenditure grouped by activities and individuals:
\begin{align*}
    B^s(\b a)&=\sum_{i=1}^n(\b a_i^s-\hb a_i^s)^2\text{ is the total expenditure on activity }s,\\
    B_j(\b a)&=\sum_{t=1}^k(\b a_j^t-\hb a_j^t)^2\text{ is the total expenditure on individual }j.
\end{align*}
By definition,  $\sum_{s=1}^{k}B^s(\b a)=\sum_{j=1}^{n} B_j(\b a)=C.$
\begin{corollary} \label{co-1} Suppose Assumptions \ref{ass-1} and \ref{ass-2} hold.\\
(a) Across-layer allocation. For any activities \(s,t\) with \(u_t\neq 0\),  \[\lim_{C\to\infty}\frac{B^s(\b a^*)}{C}=u_s^2,\text{ and }\lim_{C\to\infty}\frac{B^s(\b a^*)}{B^t(\b a^*)}=\left(\frac{u_s}{u_t}\right)^2.\]
(b) Within-layer allocation. For any individuals \(i,j\) with \(v_j\neq0\), \[\lim_{C\to\infty}\frac{B_i(\b a^*)}{C}=v_i^2,\text{ and }\lim_{C\to\infty}\frac{B_i(\b a^*)}{B_j(\b a^*)}=\left(\frac{v_i}{v_j}\right)^2.\]
\end{corollary}
Corollary \ref{co-1} follows from Theorem \ref{th-1}(b), and shows that the planner's expenditure in each activity or individual only depends on the entries of the corresponding eigenvectors. \cite{ggg} established Corollary \ref{co-1}(b) for a single activity network. Here we extend the result to the multi-activity setting, and show that budget allocation remains proportional to the squares of the entries of the principal components. 

\begin{proposition}\label{pr-3}
Suppose all activities are pairwise complements, that is, \(\beta_{st}\leq0\) for all \(s,t\). Then there exists \(\b u\in E_{min}(\b \Phi)\) such that \(u_s\geq0\) for all \(s\).
\end{proposition}

Proposition \ref{pr-3} shows that when activities are complements, for each agent, the planner can choose to adjust marginal utilities in the same direction across activities. This allows the planner to  exploit the complementarities across activities, since agents obtain maximum utility when his activity levels are all positive or all negative.

\begin{proposition}\label{pr-4}
Suppose all activities are pairwise substitutes, that is,  \(\beta_{st}\geq0\) for all \(s,t\), with strict inequality for some \(s,t\). Then for all \(\b u\in E_{min}(\b \Phi)\), there exists \(s,t\) such that \(u_s>0\) and \(u_t<0\).
\end{proposition}

Proposition \ref{pr-4} shows a contrasting result for the case of substitute activities. Now, the planner will instead adjust the marginal utilities in different directions across activities to maximize welfare. These two propositions together highlight a significant difference between complementary and substitute activities.

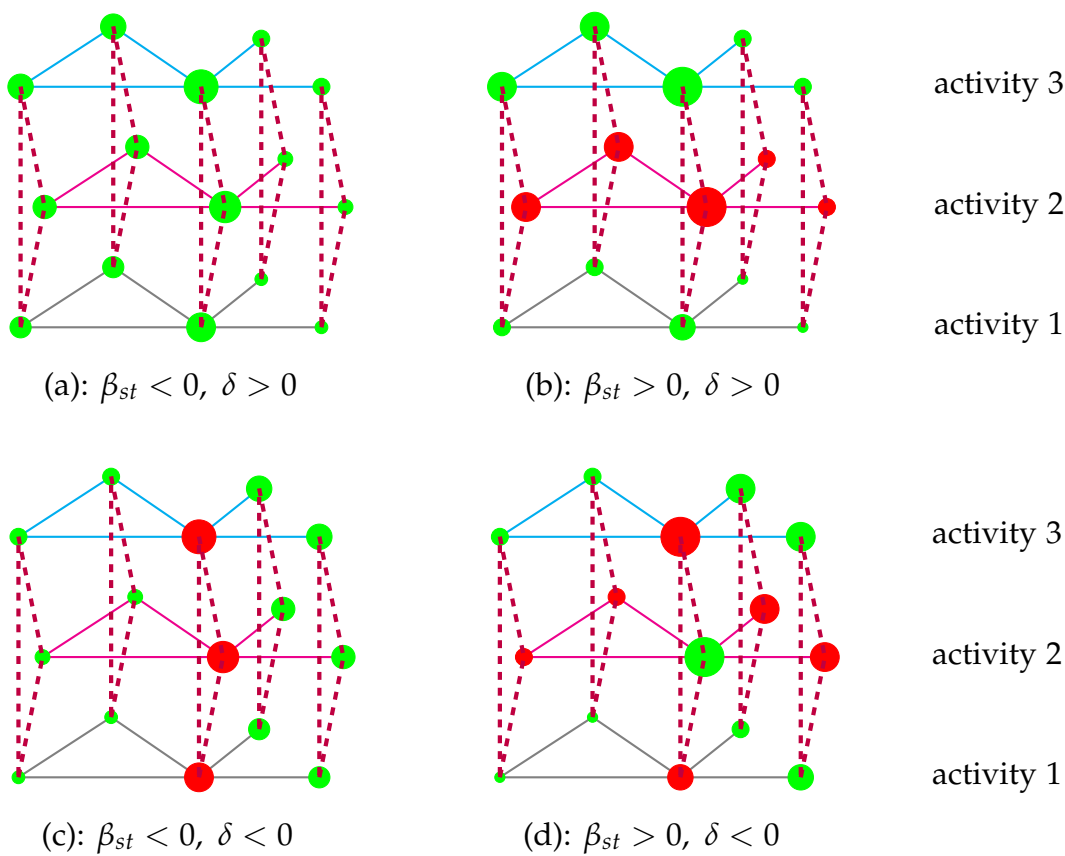
\begin{figure}[htp]
\begin{center}
\begin{tikzpicture}[scale=1.6]
        \draw[gray, thick]  (1,0) -- (0,0) -- (-0.73,0.5) -- (-1.5, 0)-- (0,0);  
          \draw[gray, thick] (0,0) -- (0.5, 0.4);  
        \draw[fill,green] (0.5,0.4) circle[radius=2.5pt];
        \draw[fill,red] (0,0) circle[radius=3.4pt];
        \draw[fill,green] (-0.73,0.5) circle[radius=1.5pt];
        \draw[fill,green] (-1.5, 0) circle[radius=1.5pt];
        \draw[fill,green] (1,0)      circle[radius=2.5pt];
        
        \begin{scope}[shift={(0.2,0)}]        
                \draw[magenta, thick]  (1,0+1) -- (0,0+1) -- (-0.73,0.5+1) -- (-1.5, 0+1)-- (0,0+1);
                        \draw[magenta, thick] (0,0+1) -- (0.5, 0.4+1);  
        \draw[fill,green] (0.5,0.4+1) circle[radius=2.75pt];
        \draw[fill,red] (0,0+1) circle[radius=3.7pt];
        \draw[fill,green] (-0.73,0.5+1) circle[radius=1.75pt];
        \draw[fill,green] (-1.5, 0+1) circle[radius=1.75pt];
        \draw[fill,green] (1,0+1)      circle[radius=2.75pt];
        \end{scope}
        
            \draw[cyan, thick]  (1,0+2) -- (0,0+2) -- (-0.73,0.5+2) -- (-1.5, 0+2)-- (0,0+2);
                    \draw[cyan,  thick] (0,0+2) -- (0.5, 0.4+2);  
           \draw[fill,green] (0.5,0.4+2) circle[radius=3pt];
            \draw[fill,red] (0,0+2) circle[radius=4pt];
        \draw[fill,green] (-0.73,0.5+2) circle[radius=2pt];
        \draw[fill,green] (-1.5, 0+2) circle[radius=2pt];
        \draw[fill,green] (1,0+2)      circle[radius=3pt];
        
           \draw[purple, dashed, ultra thick] (1,0) -- (1.2,1) --(1,2)--(1,0);
         \draw[purple, dashed, ultra thick] (0,0) -- (0.2,1) --(0,2)--(0,0);
          \draw[purple, dashed, ultra thick] (-1.5,0) -- (-1.3,1) --(-1.5, 2)--(-1.5,0);
           \draw[purple, dashed, ultra thick] (-0.73,0.5) -- (-0.53,1.5) --(-0.73,2.5)--(-0.73,0.5);
            \draw[purple, dashed, ultra thick] (0.5,0.4) -- (0.7,1.4) --(0.5,2.4)--(0.5,0.4);

    \begin{scope}[shift={(0.5,0)}]

        \draw[gray, thick]  (1+3.5,0) -- (0+3.5,0) -- (-0.73+3.5,0.5) -- (-1.5+3.5, 0)-- (0+3.5,0);  
          \draw[gray, thick] (0+3.5,0) -- (0.5+3.5, 0.4);  
        \draw[fill,green] (0.5+3.5,0.4) circle[radius=2pt];
        \draw[fill,red] (0+3.5,0) circle[radius=3pt];
        \draw[fill,green] (-0.73+3.5,0.5) circle[radius=1.2pt];
        \draw[fill,green] (-1.5+3.5, 0) circle[radius=1.2pt];
        \draw[fill,green] (1+3.5,0)      circle[radius=3pt];
        \begin{scope}[shift={(0.2,0)}]        
                \draw[magenta, thick]  (1+3.5,0+1) -- (0+3.5,0+1) -- (-0.73+3.5,0.5+1) -- (-1.5+3.5, 0+1)-- (0+3.5,0+1);
                        \draw[magenta, thick] (0+3.5,0+1) -- (0.5+3.5, 0.4+1);  
        \draw[fill,red] (0.5+3.5,0.4+1) circle[radius=3.4pt];
        \draw[fill,green] (0+3.5,0+1) circle[radius=4.6pt];
        \draw[fill,red] (-0.73+3.5,0.5+1) circle[radius=2pt];
        \draw[fill,red] (-1.5+3.5, 0+1) circle[radius=2pt];
        \draw[fill,red] (1+3.5,0+1)      circle[radius=3.4pt];
        \end{scope}
        
            \draw[cyan, thick]  (1+3.5,0+2) -- (0+3.5,0+2) -- (-0.73+3.5,0.5+2) -- (-1.5+3.5, 0+2)-- (0+3.5,0+2);
                    \draw[cyan,  thick] (0+3.5,0+2) -- (0.5+3.5, 0.4+2);  
        \draw[fill,green] (0.5+3.5,0.4+2) circle[radius=3.4pt];
        \draw[fill,red] (0+3.5,0+2) circle[radius=4.6pt];
        \draw[fill,green] (-0.73+3.5,0.5+2) circle[radius=2pt];
        \draw[fill,green] (-1.5+3.5, 0+2) circle[radius=2pt];
        \draw[fill,green] (1+3.5,0+2)      circle[radius=3.4pt];
        \begin{scope}[shift={(3.5,0)}]        
           \draw[purple, dashed, ultra thick] (1,0) -- (1.2,1) --(1,2)--(1,0);
         \draw[purple, dashed, ultra thick] (0,0) -- (0.2,1) --(0,2)--(0,0);
          \draw[purple, dashed, ultra thick] (-1.5,0) -- (-1.3,1) --(-1.5, 2)--(-1.5,0);
           \draw[purple, dashed, ultra thick] (-0.73,0.5) -- (-0.53,1.5) --(-0.73,2.5)--(-0.73,0.5);
            \draw[purple, dashed, ultra thick] (0.5,0.4) -- (0.7,1.4) --(0.5,2.4)--(0.5,0.4);
        \end{scope}
        \node[right] at (5.5,0) {activity $1$};
\node[right] at (5.5,1) {activity $2$};
\node[right] at (5.5,2) {activity $3$};
\node[right] at (-1.9,-0.5) {(c): $\beta_{st}<0,\ \delta<0$};
\node[right] at (2.1,-0.5) {(d): $\beta_{st}>0,\ \delta<0$};
\end{scope}
        \end{tikzpicture}
\end{center}
\caption{Optimal intervention for multi-activity networks}
\label{fig-1}
\end{figure}

\begin{example}\label{ex-1} We illustrate the above results on a network over five agents, and first consider the base case of having only one activity (Fig. \ref{fig:1}). A green node represents an intervention in the positive direction, while a red node represents an intervention in the negative direction. The size of the node depicts the magnitude of the intervention.\footnote{The adjacency matrix \(\mathbf G=\begin{pmatrix}0&1&1&0&0\\1&0&1&0&0\\1&1&0&1&1\\0&0&1&0&0\\0&0&1&0&0\end{pmatrix}\) has five distinct eigenvalues: \(\lambda_1=\lambda_{\max}=2.343, \lambda_2=0.471,\lambda_3=0,\lambda_4=-1,\lambda_5=\lambda_{\min}=-1.814\), so Assumption \ref{ass-2} holds.}

    \begin{figure}[htp]
\begin{center}
\begin{tikzpicture}[scale=1.6]
        \draw[gray, thick]  (1,0) -- (0,0) -- (-0.73,0.5) -- (-1.5, 0)-- (0,0);  
          \draw[gray, thick] (0,0) -- (0.5, 0.4);  
        \draw[fill,green] (0.5,0.4) circle[radius=1.5pt];
        \draw[fill,green] (0,0) circle[radius=3.4pt];
        \draw[fill,green] (-0.73,0.5) circle[radius=2.5pt];
        \draw[fill,green] (-1.5, 0) circle[radius=2.5pt];
        \draw[fill,green] (1,0)      circle[radius=1.5pt];

        \begin{scope}[shift={(4,0)}]
        \draw[gray, thick]  (1,0) -- (0,0) -- (-0.73,0.5) -- (-1.5, 0)-- (0,0);  
          \draw[gray, thick] (0,0) -- (0.5, 0.4);  
        \draw[fill,green] (0.5,0.4) circle[radius=2.5pt];
        \draw[fill,red] (0,0) circle[radius=3.4pt];
        \draw[fill,green] (-0.73,0.5) circle[radius=1.5pt];
        \draw[fill,green] (-1.5, 0) circle[radius=1.5pt];
        \draw[fill,green] (1,0)      circle[radius=2.5pt];
        \end{scope}
\node[right] at (-0.8,-0.5) {(a): $\delta>0$};
\node[right] at (3.2,-0.5) {(b): $\delta<0$};
        \end{tikzpicture}
\end{center}
\caption{Optimal intervention for single-activity networks}
\end{figure}
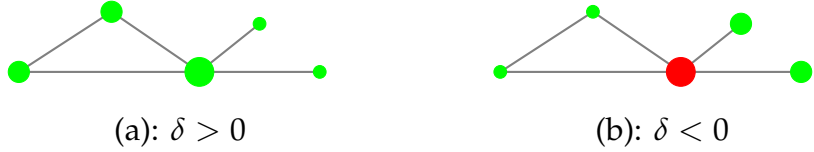\label{fig:1}
\end{example}
These illustrate the contrasting effects of the network spillovers on the optimal intervention. As shown in \cite{ggg}, when \(\delta>0\), the planner follows the first principal component of the adjacency matrix and intervenes in the positive direction. However, when \(\delta<0\), the planner follows the last principal component of the adjacency matrix, and increases the marginal utilities of some agents, while decreasing those of others.
\begin{example}
We continue with the previous example, increasing the number of activities that the agents participate in to three.
The four figures below showcase all possible permutations of complementary or substitute activities and spillovers. Figures \ref{fig-1}(a) and \ref{fig-1}(b) extend the case of strategically complementary spillovers in Example \ref{ex-1}(a) to a three activity network, while Figures \ref{fig-1}(c) and \ref{fig-1}(d) are the corresponding extension of Example \ref{ex-1}(b), with strategically substitutive spillovers.\footnote{We choose the strategic interdependence matrix \(\mathbf\Phi=\pm\begin{pmatrix}0&0.2&0.3\\0.3&0&0.4\\0.3&0.4&0\end{pmatrix}\) with eigenvalues \(\lambda_1=0.608,\lambda_2=-0.189,\lambda_3=-0.419\) in the positive case, and \(\lambda_1=0.419,\lambda_2=0.189,\lambda_3=-0.608\) in the negative case, so Assumption \ref{ass-2} holds.}\end{example}

    \begin{figure}[htp]
\begin{center}
\begin{tikzpicture}[scale=1.6]
        \draw[gray, thick]  (1,0) -- (0,0) -- (-0.73,0.5) -- (-1.5, 0)-- (0,0);  
          \draw[gray, thick] (0,0) -- (0.5, 0.4);  
        \draw[fill,green] (0.5,0.4) circle[radius=1.5pt];
        \draw[fill,green] (0,0) circle[radius=3.4pt];
        \draw[fill,green] (-0.73,0.5) circle[radius=2.5pt];
        \draw[fill,green] (-1.5, 0) circle[radius=2.5pt];
        \draw[fill,green] (1,0)      circle[radius=1.5pt];
        
        \begin{scope}[shift={(0.2,0)}]
                \draw[magenta, thick]  (1,0+1) -- (0,0+1) -- (-0.73,0.5+1) -- (-1.5, 0+1)-- (0,0+1);
                        \draw[magenta, thick] (0,0+1) -- (0.5, 0.4+1);  
        \draw[fill,green] (0.5,0.4+1) circle[radius=1.75pt];
        \draw[fill,green] (0,0+1) circle[radius=3.7pt];
        \draw[fill,green] (-0.73,0.5+1) circle[radius=2.75pt];
        \draw[fill,green] (-1.5, 0+1) circle[radius=2.75pt];
        \draw[fill,green] (1,0+1)      circle[radius=1.75pt];
        \end{scope}
        
            \draw[cyan, thick]  (1,0+2) -- (0,0+2) -- (-0.73,0.5+2) -- (-1.5, 0+2)-- (0,0+2);
                    \draw[cyan,  thick] (0,0+2) -- (0.5, 0.4+2);  
           \draw[fill,green] (0.5,0.4+2) circle[radius=2pt];
            \draw[fill,green] (0,0+2) circle[radius=4pt];
        \draw[fill,green] (-0.73,0.5+2) circle[radius=3pt];
        \draw[fill,green] (-1.5, 0+2) circle[radius=3pt];
        \draw[fill,green] (1,0+2)      circle[radius=2pt];
        
           \draw[purple, dashed, ultra thick] (1,0) -- (1.2,1) --(1,2)--(1,0);
         \draw[purple, dashed, ultra thick] (0,0) -- (0.2,1) --(0,2)--(0,0);
          \draw[purple, dashed, ultra thick] (-1.5,0) -- (-1.3,1) --(-1.5, 2)--(-1.5,0);
           \draw[purple, dashed, ultra thick] (-0.73,0.5) -- (-0.53,1.5) --(-0.73,2.5)--(-0.73,0.5);
            \draw[purple, dashed, ultra thick] (0.5,0.4) -- (0.7,1.4) --(0.5,2.4)--(0.5,0.4);

    \begin{scope}[shift={(0.5,0)}]

        \draw[gray, thick]  (1+3.5,0) -- (0+3.5,0) -- (-0.73+3.5,0.5) -- (-1.5+3.5, 0)-- (0+3.5,0);  
          \draw[gray, thick] (0+3.5,0) -- (0.5+3.5, 0.4);  
        \draw[fill,green] (0.5+3.5,0.4) circle[radius=1.2pt];
        \draw[fill,green] (0+3.5,0) circle[radius=3pt];
        \draw[fill,green] (-0.73+3.5,0.5) circle[radius=2pt];
        \draw[fill,green] (-1.5+3.5, 0) circle[radius=2pt];
        \draw[fill,green] (1+3.5,0)      circle[radius=1.2pt];
        
        \begin{scope}[shift={(0.2,0)}]
                \draw[magenta, thick]  (1+3.5,0+1) -- (0+3.5,0+1) -- (-0.73+3.5,0.5+1) -- (-1.5+3.5, 0+1)-- (0+3.5,0+1);
                        \draw[magenta, thick] (0+3.5,0+1) -- (0.5+3.5, 0.4+1);  
        \draw[fill,red] (0.5+3.5,0.4+1) circle[radius=2pt];
        \draw[fill,red] (0+3.5,0+1) circle[radius=4.6pt];
        \draw[fill,red] (-0.73+3.5,0.5+1) circle[radius=3.4pt];
        \draw[fill,red] (-1.5+3.5, 0+1) circle[radius=3.4pt];
        \draw[fill,red] (1+3.5,0+1)      circle[radius=2pt];
        \end{scope}
        
            \draw[cyan, thick]  (1+3.5,0+2) -- (0+3.5,0+2) -- (-0.73+3.5,0.5+2) -- (-1.5+3.5, 0+2)-- (0+3.5,0+2);
                    \draw[cyan,  thick] (0+3.5,0+2) -- (0.5+3.5, 0.4+2);  
        \draw[fill,green] (0.5+3.5,0.4+2) circle[radius=2pt];
        \draw[fill,green] (0+3.5,0+2) circle[radius=4.6pt];
        \draw[fill,green] (-0.73+3.5,0.5+2) circle[radius=3.4pt];
        \draw[fill,green] (-1.5+3.5, 0+2) circle[radius=3.4pt];
        \draw[fill,green] (1+3.5,0+2)      circle[radius=2pt];
        
        \begin{scope}[shift={(3.5,0)}]
           \draw[purple, dashed, ultra thick] (1,0) -- (1.2,1) --(1,2)--(1,0);
         \draw[purple, dashed, ultra thick] (0,0) -- (0.2,1) --(0,2)--(0,0);
          \draw[purple, dashed, ultra thick] (-1.5,0) -- (-1.3,1) --(-1.5, 2)--(-1.5,0);
           \draw[purple, dashed, ultra thick] (-0.73,0.5) -- (-0.53,1.5) --(-0.73,2.5)--(-0.73,0.5);
            \draw[purple, dashed, ultra thick] (0.5,0.4) -- (0.7,1.4) --(0.5,2.4)--(0.5,0.4);
                    \end{scope}
        \node[right] at (5.5,0) {activity $1$};

\node[right] at (5.5,1) {activity $2$};
\node[right] at (5.5,2) {activity $3$};
\node[right] at (-1.9,-0.5) {(a): $\beta_{st}<0,\ \delta>0$};
\node[right] at (2.1,-0.5) {(b): $\beta_{st}>0,\ \delta>0$};
        \end{scope}
        \end{tikzpicture}
\end{center}
\end{figure}

Figures \ref{fig-1}(a) and \ref{fig-1}(c) correspond to pairwise complementary activities. As in Proposition \ref{pr-3}, the planner interventions in the same direction across activities. This is particularly evident in Figure \ref{fig-1}(c), where the planner intervenes negatively for the central agent for all activities. 

On the other hand, Figures \ref{fig-1}(b) and \ref{fig-1}(d) feature pairwise substitute activities to demonstrate Proposition \ref{pr-4}. The last principal component chosen is only negative in the second entry, leading to the inversion of the signs in activity two.

\subsection{Comparative Statics}

\begin{proposition}\label{pr-5}
Suppose Assumption \ref{ass-1} holds, and we have matrices \(\b\Phi=\{\beta_{st}\},\ \b\Phi'=\{\beta'_{st}\}\) satisfying \(
\beta_{st}'\leq\beta_{st}\leq 0\) for all \(s,t\). Then \(\Gamma(\b \Phi'',\delta\b G)\geq \Gamma(\b \Phi',\delta\b G)\).
\end{proposition}

Proposition \ref{pr-5} provides an easier method of comparing the marginal returns of the planner's budget. Instead of having to evaluate the minimal eigenvalue of the strategic interdependence matrices as in Proposition \ref{pr-2}, we show that as long as the activities are all pairwise complements ($\beta_{ij}\leq 0$ for all \(i,j\)), any increase in the magnitude of the strategic interactions will result in an increase in welfare.

However, this monotonicity result does not apply in the case of substitute activities, which we demonstrate in the below example.
\begin{example}
Suppose \(\lmax(\delta\b G)=0.2\), and let
\[\b \Phi=\begin{pmatrix}0&0.2&0.3\\0.2&0&\beta_{23}\\0.3&\beta_{23}&0\end{pmatrix}.\]
We verify that Assumption \ref{ass-1} holds for \(\beta_{23}\in\{0.1,0.2,0.3\}\), and obtain the following table.
\begin{center}
\begin{tabular}{|c|c|c|c|}
    \hline
    \(\beta_{23}\)&0.1&0.2&0.3\\\hline
    $\Gamma$ &1.48&1.4&1.56\\\hline
\end{tabular}
\end{center}
\end{example}

Here, the intermediate value of \(\beta_{23}=0.2\) results in a decreased effectiveness of the planner's intervention compared to a lower or higher \(\beta_{23}\). Therefore, it is in general unclear if more strategic interaction among the agents will be beneficial for the agents.

The scope of Proposition \ref{pr-5} can be widened by considering the following transformations on the betas. For any \(t\), we can replace \[\b x^t,\hb a^t,\beta_{t1},\beta_{t2},\cdots,\beta_{tk}\text{ by }-\b x^t,-\hb a^t,-\beta_{t1},-\beta_{t2},\cdots,-\beta_{tk}\]
with no effect on the equilibrium welfare and intervention. In particular, when there are three activities, these transformations allow for the division of the possible signs of the betas to two equivalence classes:\\

(A): \(\b \Phi=\begin{bmatrix}0&+&+\\+&0&+\\+&+&0\end{bmatrix}\to\begin{bmatrix}0&-&-\\-&0&+\\-&+&0\end{bmatrix}\to\begin{bmatrix}0&-&+\\-&0&-\\+&-&0\end{bmatrix}\to\begin{bmatrix}0&+&-\\+&0&-\\-&-&0\end{bmatrix}\), and\\
(B): \(\b \Phi=\begin{bmatrix}0&+&+\\+&0&-\\+&-&0\end{bmatrix}\to\begin{bmatrix}0&+&-\\+&0&+\\-&+&0\end{bmatrix}\to\begin{bmatrix}0&-&+\\-&0&+\\+&+&0\end{bmatrix}\to\begin{bmatrix}0&-&-\\-&0&-\\-&-&0\end{bmatrix}\).\\

Therefore, with three activities, any situation with exactly two positive \(\beta_{st}\) is isomorphic to the situation of having all complementary activities, and the monotonicity result in Proposition \ref{pr-5} will hold.

\subsection{Small Budgets}
We also consider the other end of the spectrum, where the planner's budget is small.
\begin{proposition}\label{pr-6}
Suppose Assumption \ref{ass-1} holds, and the original marginal utilities are not all zero. 

(a) The optimal welfare satisfies \[\lim_{C\to0}\frac{W^*(C)-W^*(0)}{\sqrt{C}}=\|\b P\hb a\|.\]
(Note that  $\b P$ is defined in \eqref{eq-P}.)

(b) The optimal intervention satisfies \[\lim_{C\to 0}\rho(\b P \hb a,\b a^*)=1.\]
\end{proposition}

\color{red}

\color{black}

Proposition \ref{pr-6} shows that when budgets are small, the original vector of marginal utilities \(\hb a\) play a crucial role in determining the optimal intervention. This is in contrast with Theorem \ref{th-1}, where the original marginal utilities are not relevant and it is the spectral properties of the matrices \(\b \Phi\) and \(\b G\) that are important instead. \\
Another key difference between small and large budgets is the growth rate of welfare. For \(C\to 0\), Proposition \ref{pr-6}(a) shows that the optimal gain in welfare is of order \(\sqrt C\), while we have seen from Theorem \ref{th-1}(a) that as \(C\to\infty\), the planner can obtain a welfare gain that is linear in \(C\). This implies that the planner experiences significant diminishing marginal returns to the size of the budget when the budget is small, but eventually the marginal returns plateaus to a positive constant \(\Gamma(\mathbf \Phi,\delta\b G)\). Therefore, while the initial intervention is the most cost-effective, an increase in the budget will still be meaningful for the planner across all budget sizes.
\section{Partial Interventions}\label{sect-4}
We turn to a modified version of the planner's problem, where the planner is only able to intervene in a subset \(\mathcal L=\{1,\cdots,l\}\subseteq \mathcal K\) of the activities available.\footnote{This can always be achieved by a relabelling of the activities.} When \(\mathcal L=\mathcal K\), we obtain the original formulation of the problem. Under such a restriction, the planner solves the optimization problem \eqref{eq-objective}, but with an additional constraint:
\begin{align*}
    \max_{\b a\in\mathbb R^{kn}} \quad &\sm[N]iU_i(\b x_i^*(\b a);\b x_{-i}^*(\b a),\b a)\stepcounter{equation}\tag{\theequation}\label{eq-3}\\
    \text{s.t.} \quad&\b x^*_i(\b a)\in\underset{\b x_i\in\mathbb R^k}{\text{argmax}}\ U_i(\b x_i;\b x_{-i}^*(\b a),\b a)\text{ for all }i\in\c N,&\text{(Agents' equilibrium)}\\
    &(\b a-\hb a)^T(\b a-\hb a)\leq C,&\text{(Budget constraint)}\\
    &\b a^s=\hb a^s\text{ for all }s\in\mathcal K\setminus\c L.&\text{(Intervention restriction)}
\end{align*}

In this section, our focus is to study the difference in the planner's decisions when planner faces a restriction in the intervention, which we parametrize by \(l\). Therefore, we extend our previous notation and denote the optimal intervention as \(\b a^*(l,C)\), and optimal welfare \(W^*(l,C)\). 
Additionally, to simplify our analysis and focus on the key issue of a restricted intervention space, we assume that the cross-activity interactions are homogeneous, so that \(\beta_{st}=\beta\) for all \(s\neq t\). 
\begin{assumption}\label{ass-3}
 In other words, we have \(\b\Phi=\beta(\b J_k-\b I_k)\).
\end{assumption}
\subsection{Analysis}
The agents' choice of \(\b x^*\) is unaffected by the restriction in the planner's intervention space, so we know from Proposition \ref{pr-1} that the planner chooses a feasible intervention \(\b a\) to maximize \(\b a^T\b P\b a\). Define \(\c H=\c K\setminus\c L\), and decompose
\[\b a=\begin{bmatrix}\b a^{\c L}\\\b a^{\c H}\end{bmatrix},\hb a=\begin{bmatrix}\hb a^{\c L}\\\hb a^{\c H}\end{bmatrix},\b P=\begin{bmatrix}\b P^{\c {LL}}&\b P^{\c {LH}}\\\b P^{\c {HL}}&\b P^{\c {HH}}\end{bmatrix}\]
in the natural way, so that \(\b a^{\c L}\) and \(\hb a^{\c L}\) are length \(\ell n\) vectors, and \(\b P^{\c {LL}}\) is a \(\ell n\times \ell n\) matrix. The restriction on the intervention means that \(\b a^{\c H}=\hb a^{\c H}\), so we can then rewrite the problem \eqref{eq-3} as 
\begin{align*}
    \max_{\b a^{\c L}\in\mathbb R^{ln}} \quad &\b a^T\b P\b a=(\b a^{\c L})^T\b P^{\c {LL}}\b a^{\c L}+2(\hb a^{\c H})^T\b P^{\c {HL}}\b a^{\c L}+(\hb a^{\c H})^T\b P^{\c {HH}}\hb a^{\c H}\stepcounter{equation}\tag{\theequation}\label{eq-4}\\
    \text{s.t.} \quad 
    &(\b a^{\c L}-\hb a^{\c L})^T(\b a^{\c L}-\hb a^{\c L})\leq C.
\end{align*}
Applying Lemma \ref{lem-1} in Appendix \ref{sect-A} to problem \eqref{eq-4}, we obtain the following result for the asymptotic welfare and intervention for large budgets.
\begin{proposition}
\label{pr-7}
Let \(W^*(l,C)\) be the solution to the optimization problem \eqref{eq-4}, and suppose that Assumption \ref{ass-1} holds.

(a) \begin{align*}\lim_{C\to\infty}\frac{W^*(l,C)}{C}&=\lmax(\b P^{\c {LL}})\\&=\begin{cases}\frac{1-\beta}{2(1-\beta-\lmax(\delta\b G))^2},&l\geq 2\text{ and } \beta\geq0;\\\frac{l(1+(k-1)\beta)}{2k(1+(k-1)\beta-\lmax(\delta\b G))^2}+\frac{(k-l)(1-\beta)}{2k(1-\beta-\lmax(\delta\b G))^2},& otherwise.\end{cases}\end{align*}

(b) \[\lim_{C\to\infty}\frac{\|\mathrm{proj}_{E_{max}(\b P^{\c {LL}})}\ (\b a^\c L)^*\|}{\|(\b a^\c L)^*\|}=1.\]
\end{proposition}

This generalises the cosine similarity result obtained in \cite{ggg} to eigenspaces of arbitrary dimension - in particular, when \(E_{max}(\b P^{\c {LL}})=\text{span}\{\b a_0\}\) is of dimension one, the ratio in Proposition \ref{pr-7} is by definition equivalent to \(\rho((\b a^\c L)^*, \b a_0)\).  We see below in Theorem \ref{th-2} that there are important cases in our multiple activity setting where the maximum eigenvalue occurs with large multiplicity, and this generalisation using the projection operator is useful.\footnote{We could, by following the guidelines in \cite{ggg}, analytically solve and implement this optimal intervention by finding the Lagrange multiplier of the budget constraint.} More practically, Lemma \ref{lem-1} also tells us that any other direction in this eigenspace will be almost optimal. Therefore, we can choose a more convenient vector lying in the eigenspace as an approximation for our intervention, and have the following:
\begin{theorem}
\label{th-2}
Suppose Assumption \ref{ass-1} holds. Let \(\b u\) be any unit vector in \(E_{max}(\delta\b G)\). Then the following interventions \(\tb a(C)\) satisfy \(\lim_{C\to\infty}\frac{W_k(\tb a(C))}{W^*_k(l,C)}=1\):\\
(i) When the activities are complements, i.e., \(\beta<0\), we can choose the intervention \[\tb a=\sqrt{\frac{C}{l}}(\b u^T,\cdots,\b u^T,\b 0^T,\cdots, \b 0^T)^T+\hb a.\]
(ii) When the activities are substitutes, i.e., \(\beta>0\),\\
(iia) If \(l>1\), we choose\[\tb a=\sqrt{\frac{C}{2}}(\b u^T,-\b u^T,\b 0^T,\cdots,\b 0^T)^T+\hb a.\]
(iib) If \(l=1\), we choose
\[\tb a=\sqrt{C}(\b u^T,\b 0^T,\cdots,\b 0^T)^T+\hb a.\]
\end{theorem}

As in Theorem \ref{th-1}(b), we construct \emph{simple} interventions that only depend on the eigenvectors of the network matrix. Under homogeneous cross-activity interactions, we find that when the activities are complements, the planner should intervene equally in all the possible activities, where the within-activity intervention remains parallel to \(\b u\in E_{max}(\delta\b G)\) as in \cite{ggg}. As the planner becomes able to intervene in more activities, he can spread out the intervention and get closer to the optimal intervention of the unrestricted case. This explains why the optimal welfare increases as the planner intervenes in more activities.

On the other hand, when the activities are substitutes, the planner can apply a partial intervention in only two activities and still obtain an almost optimal welfare. This is done by conducting an intervention parallel to \(\b u\in E_{max}(\delta\b G)\) in both activities, but in opposite directions - the marginal utility of one activity is increased, while the other is decreased. This is because in the case of substitute products, it is inefficient for agents to choose large amounts of multiple activities, so the planner focuses on encouraging one activity while discouraging the other. 

\begin{remark}
In both cases, when \(\delta>0\), the eigenvector \(\b u\in\lmax(\b G)\) is unique up to multiplication by \(-1\) by the Perron-Frobenius theorem. On the other hand, when \(\delta<0\), no such result exists. This is because the dimension of the eigenspace of \(\lmin(\b G)\) can be up to \(n-1\), such as in the case of a complete network with an equal weight on each edge. Taking these results together, we see that we can only be certain that the eigenspace \(E_{max}(\b P^{\c {LL}})\) is of dimension 1 when \(\delta>0\) and \(\beta<0\).
\end{remark}

We return to our aim of evaluating the impact of an intervention restriction on the optimal welfare. To do so, we denote by \(\widehat W\) the original welfare without any intervention, and define the welfare improvement ratio as \[\eta_k(l,C)=\frac{W^*(l,C)-\widehat W}{W^*(k,C)-\widehat W},\ 1\leq l\leq k.\]

For convenience, we also write \[\eta_k(l,\infty)=\lim_{C\to\infty}\eta_k(l,C).\]
Clearly, \(\eta_k(l,C)\) is nondecreasing in \(l\) since an increase in \(l\) expands the feasible set for the planner while leaving the objective unchanged, so \(0\leq\eta_k(l,C)\leq\eta_k(l',C)\leq\eta_k(k,C)=1\) for all \(1\leq l\leq l'\leq k\). This captures the fraction of welfare gain the planner can achieve when there is a restriction in the intervention to \(l\) activities, compared to the case where there is no such restriction. This thus provides a measure of the benefit of intervening in more activities, giving an indication of the usefulness of having more instruments of intervention for the planner.

To further simplify our expressions, we define \begin{equation}\alpha\equiv \frac{(1+(k-1)\beta)(1-\beta-\lmax(\delta\b G))^2}{(1-\beta)(1+(k-1)\beta-\lmax(\delta\b G))^2}.\label{eq-5}\end{equation}

Here \(\alpha\) depends on the number of activities, \(k\), the degree of complementarity, \(\beta\), and the network effects, \(\delta\b G\), but is independent of the intervention restriction, \(l\). Under Assumption \ref{ass-1}, we can show that when the activities are complements \((\beta<0)\) then \(\alpha>1\), when the activities are substitutes \((\beta>0)\) then \(0<\alpha<1\), and when the activities are independent \((\beta=0)\) then \(\alpha=1\). 

We have the following theorem on the behaviour of \(\eta_k\) when the budget is large.

\begin{theorem}
\label{th-3} Suppose Assumptions \ref{ass-1} and \ref{ass-3} hold.\\
(i) If the activities are substitutes, i.e., \(\beta>0\), then
\begin{equation}\eta_k(l,\infty)=\begin{cases}
\frac{1}{k}(k-1+\alpha),&l=1;\\1,&l\geq2.\end{cases}\label{eq-6}
\end{equation}
(ii) If the activities are complements, i.e., \(\beta<0\), then
\begin{equation}\eta_k(l,\infty)=
\frac{1}{k}\left(l+\frac{k-l}{\alpha}\right).\label{eq-7}
\end{equation}
\end{theorem}

Theorem \ref{th-3} follows directly from Proposition \ref{pr-7}. From equation \eqref{eq-6}, when activities are substitutes, there is a loss in welfare when the planner can only intervene in one activity, but the planner can reach an asymptotically optimal intervention as long as there are at least two activities that can be intervened in. That is, \[\eta_k(1,\infty)<\eta_k(2,\infty)=\eta_k(3,\infty)=\cdots=\eta_k(k,\infty)=1.\]Furthermore, the planner is able to achieve at least \(\frac{k-1}{k}\) of the optimal welfare even when there is only one activity available for intervention ($l=1$), so the welfare loss will be small if the agents are simultaneously participating in many activities. As a result, a social planner does not need to consider implementing a variety of intervention channels across the multiple activities when the activities are substitutes, and it suffices to focus on one or two of them.

On the other hand, when the activities are complements, Theorem \ref{th-3}(ii) shows that the planner cannot reach an optimal welfare whenever there are activities the planner is unable to intervene in. That is, \[\eta_k(1,\infty)<\eta_k(2,\infty)<\cdots<\eta_k(k,\infty)=1.\] Furthermore, we see from \eqref{eq-7} that the asymptotic welfare ratio is linear in \(l\), which implies  that the gain in relative welfare is linear in the number of restricted activities.\footnote{Formally, we obtain $\partial \eta_k(l,\infty)/\partial l=(\alpha-1)/(\alpha k)$.} This demonstrates a stark contrast between the cases where the activities are substitutes or complements. Only when the activities are complements does the planner benefit from intervention in multiple activities. This induces an increase in the agents' participation in more activities and hence improves total welfare from the positive interactions between them.

Finally, when the activities are independent, we have \(\beta=0\), thus, $\alpha=1$ and \(\eta_k(l,\infty)=1\) for all \(k\) and \(l\). The planner asymptotically reaches the same level of welfare regardless of the number of activities that can be intervened in. In this situation, the planner only cares about the relative consumption across agents for each activity, and not the relative consumption across activities.

Theorem \ref{th-3} generalizes the findings in \cite{ggg} to a setting with multiple activities. We see that the presence of interactions between the activities, as well as the restriction on the number of activities that the planner can intervene in, play a significant role in determining the effectiveness of the intervention. In particular, a 2-activity partial intervention is sufficient in the case where the activities are substitutes, but a partial intervention will not be as effective as a complete intervention when the activities are complements.

To complete our analysis, we also obtain the welfare loss for small budgets.

\begin{theorem}
\label{th-4}
Suppose Assumption \ref{ass-1} holds, and the original marginal utilities are not all zero. For all \(\beta\) and \(l\), \[\lim_{C\to0}\eta_k(l,C)=\frac{\|\b P^{\c L}\hb a\|}{\|\b P\hb a\|}.\]
In particular, if the activities are homogeneous with \(\hb a^s=\hb a^t\) for all \(s,t\in\c K\), then  \[\lim_{C\to0}\eta_k(l,C)=\sqrt{\frac{l}{k}}.\]
\end{theorem}

In contrast to the result for large \(C\), here the original marginal utility \(\hb a\) is crucial in determining the optimal welfare gain, while there is no significant dependence on the sign of \(\beta\). This echoes the result for the single activity case in \cite{ggg}. Therefore, in general, the choice of activities that allow for intervention will affect the results.

Since we have that \[\sum_{s\in\c L}\|\b P^{\{s\}}\hb a\|^2=\|\b P^{\c L}\hb a\|^2,\]
given a fixed number of activities \(l\), there exists a choice of \(\c L\) with \(|\c L|=l\) such that \(\eta_k(l,C)\geq\sqrt{\frac{l}{k}}\). This shows that as long as the planner is able to pick the activities to intervene in, there are decreasing marginal returns to the number of activities in which the planner intervenes. This is in contrast to the linear gains found in Theorem \ref{th-3}. Therefore, when the budget is small, it is important for the planner to choose to intervene in the correct activities, after which including other activities will provide a smaller welfare gain. In particular, if the activities are homogeneous, then all choices of \(\c L\) of the same size result in the same welfare, and equality always holds.

We illustrate above results on the optimal welfare for a simple network below. 

\begin{example}
 Consider a dyad network\footnote{\(\b G^{dyad}=\begin{pmatrix}0&1\\1&0\end{pmatrix}\).} over three activities, setting \(\delta=0.1\), and suppose the activities are ex ante homogeneous with \(\hb a^1=\hb a^2=\hb a^3=(2,1)^T\). Letting the number of activities that allow for interventions to vary, we obtain the following plots for optimal welfare as the budget $C$ ranges from 0 to 40:

\begin{figure}[ht!]
\begin{subfigure}{0.5\textwidth}
    \centering
   \includegraphics[width=1\linewidth]{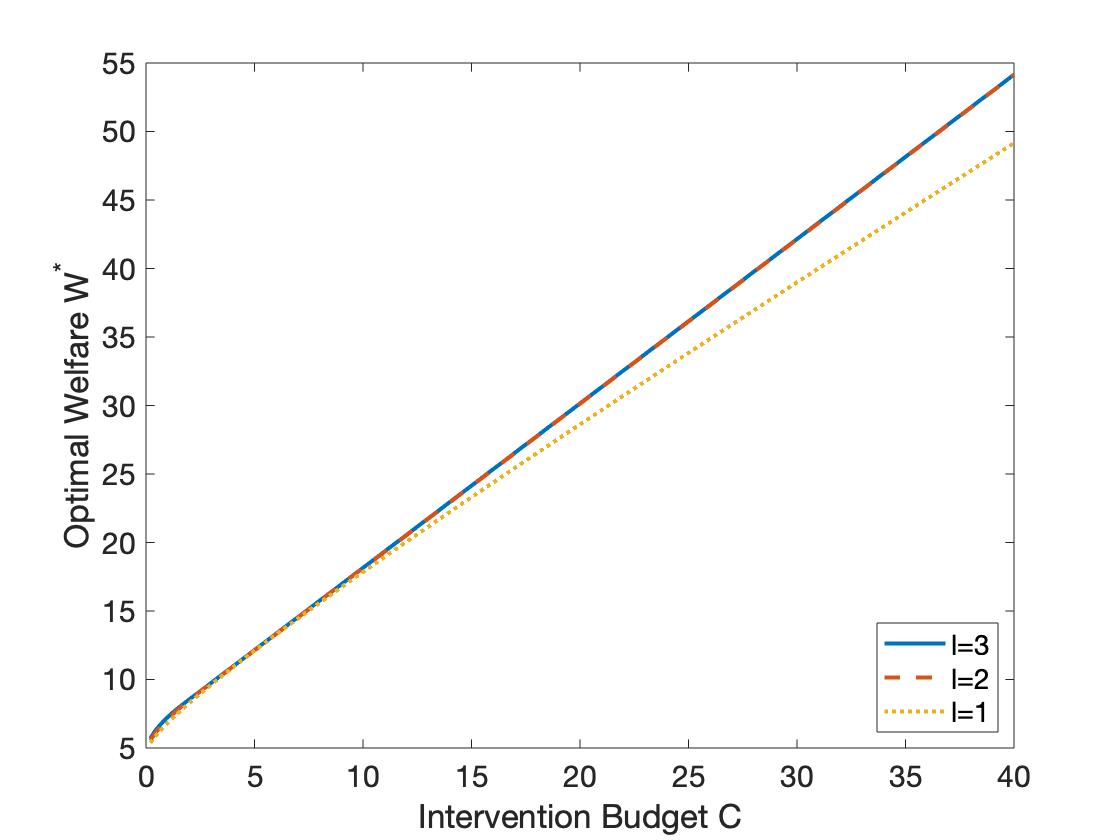}
    \caption{\(\beta=0.4\)}
    \label{Figure 4.1}
\end{subfigure}
\begin{subfigure}{0.5\textwidth}
    \centering
    \includegraphics[width=1\linewidth]{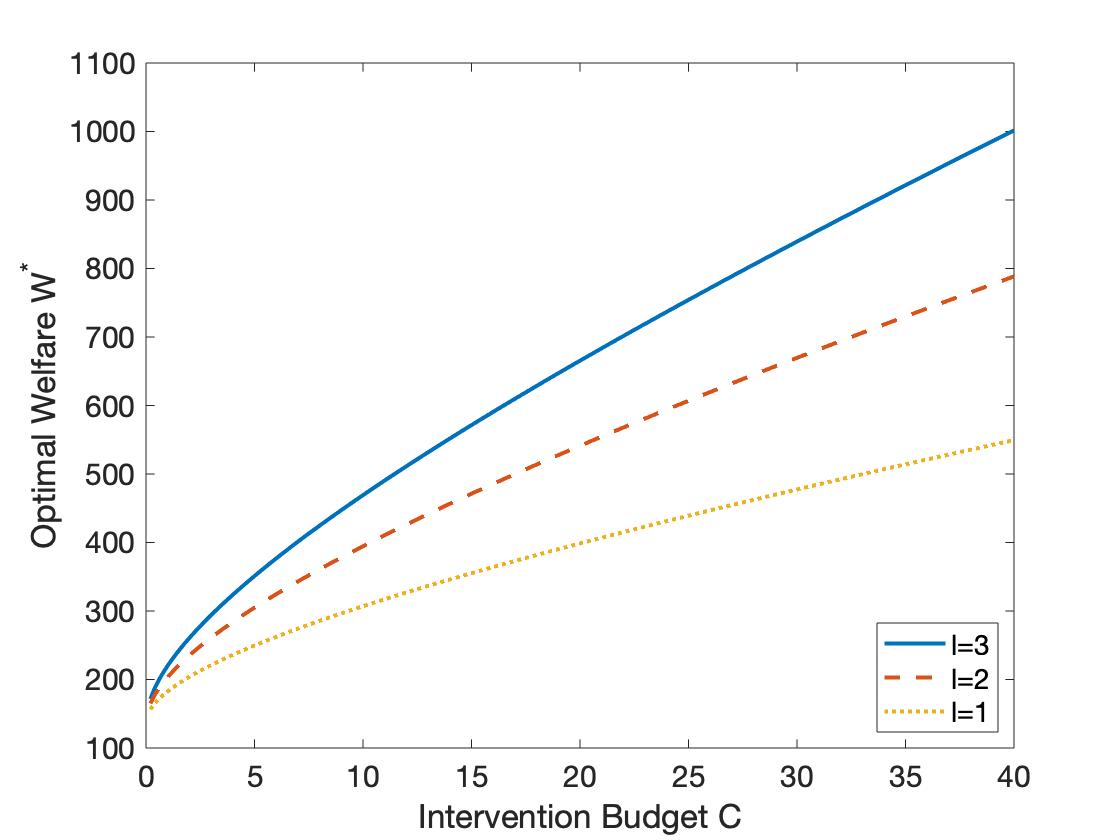}
    \caption{\(\beta=-0.4\)}
    \label{Figure 4.2}
\end{subfigure}
\newline
\caption{Optimal welfare against budget for varying intervention restrictions}
\label{fig:2}
\end{figure}
\end{example}

As we expect from Proposition \ref{pr-7}, the graphs in Figures \ref{fig:2}(a) and \ref{fig:2}(b) exhibit linear growth as \(C\) grows large. In Figure \ref{fig:2}(a), the curves for \(l=2\) and \(l=3\) are almost identical, reflecting our result in Theorem \ref{th-3} that when the activities are substitutes \((\beta>0)\), an intervention in two activities is always sufficient to obtain an almost optimal welfare. On the other hand, Figure \ref{fig:2}(b) shows the necessity of intervening in more activities in the case of complementary activities \((\beta<0)\), and the incremental effect of an additional activity is asymptotically constant. On the other hand, we see that the gap between the curves \(l=1\) and \(l=2\) is larger than the gap between \(l=2\) and \(l=3\) when \(C\) is small, as we have seen in Theorem \ref{th-4}.

\subsection{Comparative Statics}
As a counterpart to Proposition \ref{pr-5}, we examine how the normalized welfare \(\Gamma\) is affected by a change in the model parameters in this case.
\begin{proposition}\label{pr-8} Suppose Assumption \ref{ass-1} holds, and \(\Phi=\beta(\b J_k-\b I_k)\). Then for any \(1\leq l\leq k\), \\
(i) \(\Gamma(\Phi,\delta\b G,l)\) is increasing in \(\lmax(\delta\b G)\).\\
(ii) \(\Gamma(\Phi,\delta\b G,l)\) is increasing in \(|\beta|\).
\end{proposition}

Here, the homogeneity of the cross-activity interactions allows us to extend the monotonicity result to the case of substitute activities. An increase in the intensity of either the cross-activity interactions or the network spillovers will lead to an increase in the effectiveness of the intervention, as these spillovers serve to propagate the effects of the intervention throughout the network and act as a multiplier for the equilibrium welfare gain. Of further interest is the effect of these model parameters on the welfare ratio discussed in Theorem \ref{th-3}. From Theorem \ref{th-3}, we know that
\[\lim_{C\to\infty}\frac{\partial\eta_k(l,C,\b G)}{\partial\alpha}=\begin{cases}-\frac{k-l}{k\alpha^2},&\beta<0;\\\frac{1}{k},&\beta>0,\ l=1;\\0,&otherwise.\end{cases}\]
Therefore, we obtain the following comparative statics effects on the welfare ratio.

\begin{proposition}\label{pr-9}
(i) \(\eta_k(l,\infty)\) is weakly decreasing in \(\lmax(\delta\b G)\).\\
(ii) \(\eta_k(l,\infty)\) is weakly decreasing in \(|\beta|\).
\end{proposition}
Proposition \ref{pr-9} shows that the welfare ratio moves in the opposite direction of the welfare gain. Having stronger cross-activity interactions or network spillovers instead decrease the welfare ratio, making it more important for the planner to have access to intervention in all activities.

While the interpretation of a change in \(\beta\) is straightforward, we want to link changes in the network structure \(\b G\) to the spectral radius of \(\delta\b G\). We first define a partial ordering on the set of networks. Given two graphs \(\b G\) and \(\b G'\), we write \(\b G\triangleleft\b  G'\) if \(g_{ij}\leq g'_{ij}\) for all \(i,j\). In particular, \(\b G\triangleleft\b G'\) whenever \(\b G\) represents a subgraph of \(\b G'\) over the same set of agents. We have the following standard results on the largest and smallest eigenvalues of \(\b G\).
\begin{fact}
\label{fact-1}
(i) If \(\b G\triangleleft\b G'\), then \(\lmax(\b G)\leq\lmax(\b G')\).\\
(ii) If \(\b G\triangleleft\b G'\) and \(\b G'\) is bipartite, then \(\lmin(\b G)\geq\lmin(\b G')\).
\end{fact}
Using this characterization, we can restate part (i) of Proposition \ref{pr-9}.
\begin{proposition}
\label{pr-10} Let \(\b G\triangleleft\b G'\).\\
(i) If \(\delta>0\), then \(\eta_k(l,\infty,\b G)\geq\eta_k(l,\infty,\b G')\).\\
(ii) If \(\delta<0\) and \(\b G'\) is bipartite, then \(\eta_k(l,\infty,\b G)\geq\eta_k(l,\infty,\b G')\).
\end{proposition}
Therefore, increasing the connectivity of the network will decrease the welfare ratio when either \(\delta>0\), or \(\delta<0\) and \(\b G'\) is bipartite. That is, when the network spillovers are positive, a denser network would result in a greater welfare loss from a partial intervention, but the effect is in general ambiguous when the spillovers are negative.

\section{Extensions}
\label{sect-5}

\subsection{Heterogeneous Network Externalities}

In the previous sections, we have imposed that the strength of the network spillovers \(\delta\) is the same for each activity. We now relax this assumption, such that each activity \(s\) has its own corresponding coefficient \(\delta_s\). That is, each agent now receives a total utility of \begin{equation}U_i= \sum_{s\in\mathcal{K}}a_i^sx_i^s-\left(\frac{1}{2}\sum_{s\in\mathcal{K}}(x_i^s)^2+\frac{1}{2}\sum_{s\in\mathcal{K}}\sum_{\substack{t\in\mathcal{K}\\t\neq s}}\beta_{st}x_i^sx_i^t\right)+\sum_{s\in\mathcal{K}}\sum_{j\in\mathcal{N}}
\delta_sg_{ij}x_i^sx_j^s.\label{eq-h}\end{equation}
It is useful to define a new matrix of spillovers \(\b \Delta=diag(\delta_1,\cdots,\delta_k)\) to capture the varying network externalities. The consumer's equilibrium choice of actions, following the similar argument as in Proposition \ref{pr-1}, can be shown to be 
\begin{equation}
\b x^*=[\tb\Phi\otimes\b I_n-\b\Delta\otimes\b G]^{-1}\b a,
\label{eq-h-x}
\end{equation}
with the corresponding total welfare given by
\begin{equation}
 W(\b a)=\frac{1}{2}\b a^T\underbrace{[\tb \Phi\otimes\b I_n-\b \Delta\otimes\b G]^{-1}(\tb \Phi\otimes\b I_n)[\tb \Phi\otimes\b I_n-\b \Delta\otimes\b G]^{-1}}_{:=\c P}\b a.
\label{eq-h-w}
\end{equation}
Here we need to impose  the following assumption, which generalizes Assumption \ref{ass-1} to ensure the well-behavedness of model under heterogeneous \(\delta_s\).
\setcounter{assumption}{0}
\renewcommand\theassumption{\arabic{assumption}'}
\begin{assumption}\label{ass-4}
The matrix \(\tb\Phi\otimes\b I_n-\b\Delta\otimes\b G\) is positive definite.
\end{assumption}
Assumption \ref{ass-4} is equivalent to the condition that \(\tb\Phi-t\b\Delta\succ 0\) for all eigenvalues \(t\) of \(\b G\).\footnote{For symmetric matrices \(\b A\) and \(\b B\), we write \(\b A\succ\b B\) if \(\b A-\b B\) is positive definite.} Under our baseline model of homogeneous network externalities, that is, \(\delta_s=\delta\) for all \(s\), then Assumption \ref{ass-4} reduces to the condition
\(
\lmin(\tb \Phi)>\lmax(\delta\b G)
\)
 and we recover Assumption \ref{ass-1}. Also, when \(\delta_s>0\) for all \(s\), Assumption \ref{ass-4} reduces to \(\tb\Phi-\lmax(\b G)\b\Delta\succ0\), while when \(\delta_s<0\) for all \(s\), Assumption \ref{ass-4} instead reduces to \(\tb\Phi-\lmin(\b G)\b\Delta\succ0\).

To state the optimal intervention under heterogeneous \(\delta_s\), we define the matrix \[\b M(t)=\tb\Phi-2t\b\Delta+t^2\b\Delta\tb\Phi^{-1}\b\Delta,\quad t\in\mathbb R.\]
In our analysis, \(\b M(t)\) performs a similar role to \(\tb\Phi\), but takes into account the effects of \(\b\Delta\) to an extent parametrized by a scalar \(t\).

\begin{proposition}\label{pr-12}
Suppose Assumption \ref{ass-4} holds, and \(\hb a=0\).\footnote{For general \(\hb a\), we could state asymptotic optimality results of such $\b a^*$ stated in Proposition \ref{pr-12} under large $C$  in the similar spirit  to Theorem \ref{th-1}. We omit the details for brevity.}

(a) If \(\delta_s>0\) for all \(s\), then the optimal intervention satisfies \(\b a^*=\sqrt{C}\b u\otimes\b v\) for some \(\b v\in E_{max}(\b G)\) and \(\b u\in E_{min}(\b M(\lmax(\b G)))\).

(b) If \(\delta_s<0\) for all \(s\), then the optimal intervention satisfies \(\b a^*=\sqrt{C} \b u\otimes\b v\) for some \(\b v\in E_{min}(\b G)\) and \(\b u\in E_{min}(\b M(\lmin(\b G)))\).
\end{proposition}

Similarly to Theorem \ref{th-1}, Proposition \ref{pr-12} is obtained by a spectral analysis of the matrix \(\c P\), which we provide in the following Proposition:
\begin{proposition}\label{pr-11}
Let \(\{(t_i,\b v_i)\}_{i=1}^n\) be a set of orthogonal eigenpairs of \(\b G\). For each \(i\), let \(\{(\mu_{ji},\b u_{ji})\}_{j=1}^k\) be a set of orthogonal eigenpairs of the matrix \(\b M(t_i)\). Then \(\{(\mu_{ji}^{-1},\b u_{ji}\otimes\b v_i)\}\) is a set of orthogonal eigenpairs of \(\c P\).
\end{proposition}

Proposition \ref{pr-11} identifies a set \(k\times n\) orthogonal eigenpairs of \(\c P\), so they must contain all the eigenvalues of \(\c P\). By applying Lemma \ref{lem-1}, Proposition \ref{pr-12} follows by selecting the largest eigenvalue of \(\c P\) and choosing its corresponding eigenspace as the optimal intervention.

Proposition \ref{pr-12} shows that our previous results on the within-activity intervention are robust to including heterogeneity in the strength of the network spillovers. If the network generates positive spillovers for each activity, the optimal intervention will still follow the first eigenvector of \(\b G\), while if the network generates negative spillovers for each activity, the optimal intervention will follow the last eigenvector of \(\b G\). The optimal across-activity intervention no longer follows the principal components of \(\tb\Phi\), but instead can be expressed in terms of the spectral properties of a new matrix \(\b M(t)\), which captures both the cross-activity interactions \(\tb\Phi\) and the heterogeneous within-activity spillovers \(\b\Delta\). 

We provide a simple example of the effects of varying \(\b\Delta\) on the optimal across-activity interaction below.
\begin{example}
Let \(k=2, \tb\Phi=\begin{bmatrix}1&-0.3\\-0.3&1\end{bmatrix},\ \delta_1=0.1\), and \(\b G\) be an arbitrary cycle graph. The table below illustrates the optimal intervention for varying \(\delta_2\).\footnote{Note \(\lambda_{\max}(\b G)=2\) with corresponding eigenvector \( \b v=(\frac{1}{\sqrt n},\cdots,\frac{1}{\sqrt n})^T\).  For each $\delta_2$, the vector $\b u$ in the table is given by the last eigenvector  of $\b M(\lmax(\b G))=\b M(2)$ by Proposition \ref{pr-12}(a). }

\begin{table}[H]
    \centering
    \begin{tabular}{|c|c|c|c|}
        \hline\(\delta_2\) & 0.1 & 0.2 & 0.3 \\
        \hline\(\b u\)& \((0.707,0.707)^T\) & \((0.483,0.876)^T\) & \((0.390,0.921)^T\)\\
        \hline\(\frac{u_1}{u_2}\)& 1 & 0.551 & 0.423\\
        \hline
    \end{tabular}
\end{table}
\end{example}
By Proposition \ref{pr-12}, the cross-activity intervention is proportional to the eigenvector \(\b u\), so \(\frac{u_1}{u_2}\) represents the ratio of the amount of intervention between activities 1 and 2 (see Corollary \ref{co-1}). When \(\delta_2=0.1\), we have homogeneous spillovers as in our baseline model, and the planner performs an equal amount of intervention in each activity. Intuitively, as \(\delta_2\) increases, we see that the ratio \(\frac{u_1}{u_2}\) decreases, as the larger spillovers in activity 2 leads to an increase in the effectiveness of intervention, and a larger proportion of the planner's budget will be allocated to activity 2.

\subsection{Nonnegative Interventions}
We return to the baseline model \eqref{eq-objective}, but now assume that the planner is only able to conduct a nonnegative intervention. That is, we must have \(\b a^*\geq\hb a\). For instance, for a planner seeking an intervention in education, a reduction in an individual's returns to education could be seen as unethical. Therefore, it is reasonable for a planner to be bound by such a constraint in an applied setting. For simplicity, we suppose that \(\hb a=\b 0\). Applying Proposition \ref{pr-1}, we can write this problem as 
\begin{align}
    \max_{\b a\in\mathbb R^{kn}}&\qquad W(\b a)=\b a^T\b P\b a\label{eq-positive}\\\text{s.t.}&\qquad \b a^T\b a\leq C,\ \b a\geq\b 0.\nonumber
\end{align}
When \(\delta>0\) and \(\beta_{st}\leq0\) for all \(s,t\), we know from Proposition \ref{pr-3} and Example \ref{ex-1} that the optimal unconstrained intervention is already nonnegative. Therefore, the additional constraint is not binding, and does not affect the solution. However, if \(\delta<0\) or \(\beta_{st}>0\) for some \(s,t\), the optimal unconstrained intervention will be negative in some component, and is not feasible. We can still perform a spectral analysis to obtain a necessary condition:
\begin{proposition}\label{pr-positive}
Suppose \(\b a^*\) solves \eqref{eq-positive}. Let the support of \(\b a^*\) be the set of indices \(S=\{i:a_i^*>0\}\), and let \(\b a_S^*\) be the subvector of \(\b a^*\) induced by the indices in \(S\). Similarly, let \(\b P_S\) be the principal submatrix of \(\b P\) induced the indices in \(S\). Then \(\b a_S^*\in E_{max}(\b P_S)\), and \(\frac{W(\b a^*)}{C}=\lmax(\b P_S)\).
\end{proposition}
Proposition \ref{pr-positive} implies that we can obtain the optimal intervention by simply checking for each principal submatrix of \(\b P\) for a nonnegative eigenvector, and finding the maximum eigenvalue among them. While this process may seem inefficient, the following proposition suggests that we will not be able to do much better.
\begin{proposition}\label{pr-14}
The problem \eqref{eq-positive} is NP-hard.
\end{proposition}
Consequently, adding the nonnegativeness constraints in the intervention problem increases its computational complexity.

Finally, we make use of two sample networks from \cite{ggg}, to illustrate the optimal nonnegative intervention in the single activity case, i.e., $k=1$. Both of these graphs can be checked to be 3-regular.

\begin{example} \label{ex-6} The following diagrams illustrate a representative optimal intervention\footnote{Multiple optimal interventions exist due to the symmetries of the graph, but they are equivalent up to a graph automorphism.} for \(\delta=-0.05\) and \(\delta=-0.2\) (Fig. \ref{fig-4}). Nodes coloured black receive zero intervention, while red nodes are labelled by their normalized optimal intervention, \(\frac{\b a^*}{\sqrt C}\).
\begin{figure}[htp]
\begin{center}
\begin{tikzpicture}[r/.style={circle, fill=black, inner sep=0pt, minimum size=2mm}]
        \node[r](1) at (0,2) {};
        \node[r,label=left:0.507,fill=red](2) at (0,0) {};
        \node[r](3) at (1,3){};
        \node[r](4) at (1,2){};
        \node[r,label=0.595,fill=red](5) at (2,1){};
        \node[r](6) at (2,0){};
        \node[r,label=0.371,fill=red](7) at (3,3){};
        \node[r](8) at (3,2){};
        \node[r](9) at (4,2){};
        \node[r,label=right:0.501,fill=red](10) at (4,0){};
        \draw (1)--(2)--(6)--(10)--(9)--(7)--(3)--(1)--(5)--(9);
        \draw (3)--(8)--(4)--(7);
        \draw (2)--(4);
        \draw (8)--(10);
        \draw (5)--(6);

        \begin{scope}[shift={(0,-5.5)}]
        \node[r](1) at (0,2) {};
        \node[r,label=left:0.518,fill=red](2) at (0,0) {};
        \node[r,label=0.180,fill=red](3) at (1,3){};
        \node[r](4) at (1,2){};
        \node[r,label=0.595,fill=red](5) at (2,1){};
        \node[r](6) at (2,0){};
        \node[r,label=0.180,fill=red](7) at (3,3){};
        \node[r](8) at (3,2){};
        \node[r](9) at (4,2){};
        \node[r,label=right:0.518,fill=red](10) at (4,0){};
        \draw (1)--(2)--(6)--(10)--(9)--(7)--(3)--(1)--(5)--(9);
        \draw (3)--(8)--(4)--(7);
        \draw (2)--(4);
        \draw (8)--(10);
        \draw (5)--(6);
        \end{scope}
        \begin{scope}[shift={(6,0.5)}]
        \node[r](1) at (0,2) {};
        \node[r](2) at (0,0) {};
        \node[r,label=0.602,fill=red](3) at (1,2){};
        \node[r,label=below:0.602,fill=red](4) at (1,0){};
        \node[r,label=85:0.469](5) at (2,1){};
        \node[r,fill=red](6) at (3,1){};
        \node[r](7) at (4,2){};
        \node[r](8) at (4,0){};
        \node[r,label=0.234,fill=red](9) at (5,2){};
        \node[r](10) at (5,0){};
        \draw (1)--(3)--(2)--(4)--(1)--(2);
        \draw (3)--(5)--(4);
        \draw (5)--(6);
        \draw (7)--(6)--(8);
        \draw (9)--(7)--(10)--(8)--(9)--(10);
        \end{scope}
        
        \begin{scope}[shift={(6,-5)}]
        \node[r](1) at (0,2) {};
        \node[r](2) at (0,0) {};
        \node[r,label=0.548,fill=red](3) at (1,2){};
        \node[r,label=below:0.548,fill=red](4) at (1,0){};
        \node[r,label=85:0.570](5) at (2,1){};
        \node[r,fill=red](6) at (3,1){};
        \node[r](7) at (4,2){};
        \node[r](8) at (4,0){};
        \node[r,label=0.193,fill=red](9) at (5,2){};
        \node[r,label=below:0.193,fill=red](10) at (5,0){};
        \draw (1)--(3)--(2)--(4)--(1)--(2);
        \draw (3)--(5)--(4);
        \draw (5)--(6);
        \draw (7)--(6)--(8);
        \draw (9)--(7)--(10)--(8)--(9)--(10);
        \end{scope}     
        \node[right] at (3.5,-1) {(a): $\delta=-0.05$};
        \node[right] at (3.5,-6.5) {(b): $\delta=-0.2$};
        \end{tikzpicture}
\end{center}
\caption{Optimal nonnegative interventions} \label{fig-4}
\end{figure}

Observe that when \(|\delta|\) is small, the nodes with positive intervention forms an independent set, but this is not the case when \(|\delta|\) is large. Recall from Proposition \ref{pr-1} (or see \cite{bcz}) that for single-activities, we can simplify the expression for welfare as \(2\b a^T\b P\b a=\b a^T[\b I-\delta\b G]^{-2}\b a\), so a Taylor expansion at \(\delta=0\) gives \[2\b a^T\b P\b a\approx\b a^T\b a+2\delta\b a^T\b G\b a.\]
Recall  $\b a^T\b a= C$.
Intuitively, for \(\delta\) sufficiently close to zero, we maximize the linear term \(2\delta\b a^T\b G\b a\). Since \(\delta<0\),  it suffices to minimize \(\b a^T\b G\b a=\sum_{i,j} g_{ij} a_i a_j\), which equals zero if and only if the support of \(\b a\) must be an independent set of $\b G$. See panel (a) of the above figure.   On the other hand, when \(|\delta|\) is large, the higher order terms in the Taylor expansion\footnote{$2\b a^T\b P\b a=\b a^T\b a+2\delta\b a^T\b G\b a+3\delta^2\b a^T\b G^2\b a+\cdots.$} become more significant, and we observe a trade-off between minimizing the odd-length paths which decrease total welfare, and maximizing the even-length paths which increase welfare. We expect similar results to hold for other graphs as well.
\end{example}

\subsection{An Alternative Intervention Model - Pricing}
Besides welfare-maximizing interventions, we show that our model can also find applications in other contexts. Here, we consider a model where a monopolist provides multiple goods to a network of consumers, as a generalization of \cite{cyz2}. We assume that the firm faces constant marginal costs, which we denote by the vector \(\b c=(c_i^1,\cdots,c_n^k)^T\), where \(c^s_i\) represents the marginal cost of producing good \(s\) to consumer \(i\). Furthermore, we assume that \(c_i^s>a_i^s\) for all \(s,i\) to ensure that the firm can obtain positive profits. The firm then chooses a price vector \(\b p=(p_1^1,\cdots,p_n^k)^T\), where \(p_i^s\) represents the price of good \(s\) to consumer \(i\). Consumer \(i\) thus incurs a total cost of \(\sum_{s=1}^k p_i^sx_i^s\) from purchasing the quantity \(\b x_i\), resulting in a total utility of \[U_i= \sum_{s\in\mathcal{K}}(a_i^s-p_i^s)x_i^s-\left(\frac{1}{2}\sum_{s\in\mathcal{K}}(x_i^s)^2+\frac{1}{2}\sum_{s\in\mathcal{K}}\sum_{\substack{t\in\mathcal{K}\\t\neq s}}\beta_{st}x_i^sx_i^t\right)+\sum_{s\in\mathcal{K}}\sum_{j\in\mathcal{N}}
\delta_sg_{ij}x_i^sx_j^.\]

For a given price vector $\b p$, the equilibrium demand, using \eqref{eq-h-x},  is given by \[\b x^*=[\tb\Phi\otimes\b I_n-\b\Delta\otimes\b G]^{-1}(\b a-\b p).\]
The firm thus solves the profit-maximizing problem \begin{equation}\label{eq-firm}\max_{\b p\in\mathbb R^{kn}}\ \pi=(\b p-\b c)^T\b x^*=(\b p-\b c)^T[\tb\Phi\otimes\b I_n-\b\Delta\otimes\b G]^{-1}(\b a-\b p)\end{equation}

\begin{proposition}\label{pr-13}
Suppose Assumption \ref{ass-4} holds. The solution to \eqref{eq-firm} is \[\b p^*=\frac{\b a+\b c}{2},\]
with a corresponding maximal monopoly profit \[\pi^*=\frac{1}{4}(\b a-\b c)^T[\tb\Phi\otimes\b I_n-\b \Delta\otimes\b G]^{-1}(\b a-\b c).\]
\end{proposition}

That is, the firm sets prices equal to the average of the intrinsic marginal utilities of the consumers and the marginal costs of production. The network structure and cross-activity interactions are irrelevant in determining optimal pricing, and only contributes to the equilibrium quantities and profit. This network-independent pricing result generalizes the optimal  pricing problem for a single good (see \cite{cbo} and \cite{bloch2013pricing}) to multiple products, and we find that similar results hold in this setting. We admit that the network independence result here is due to certain special features of the model (such as linear demand and no competition). See \cite{Bloch2015} for a survey of targeted pricing in social networks and \cite{cyz2,O2p2022} for settings with competitive pricing.\footnote{Relatedly, \cite{ggg3} study taxation in a network market under an oligopolistic setting, and find that optimal taxes and surplus can also be described in terms of the spectral properties of the network.}

\section{Conclusion}\label{sect-6}
We analyse the problem of a welfare-maximizing intervention on a heterogeneous multiplex network using principal component analysis. By solving a general quadratic programming problem, we show when the planner's budget is large, the optimal intervention and welfare can be simply described by the spectral properties of two matrices, one representing the within network spillovers among the agents, and the other representing the strategic interdependence between activities. We study the effect of this interdependence on welfare, and find that when activities are pairwise complements, the marginal welfare gained from the planner's budget increases with the degree of complementarity. Furthermore, we establish a class of strategic interdependencies that exhibit the same property. We also consider the problem for small budgets, and show that in contrast, the optimal intervention becomes dependent on the original utilities and not the spectral decomposition. However, our paper only considers the situation where underlying network is the same across each activity, and future research may be able to weaken this strong assumption.

As an application of our model, we have also studied a related problem where the planner is unable to intervene in all activities. We analyse the optimal welfare under such a constraint, and find that there are significant differences in the results depending on whether activities are substitutes or complements, as well as the size of the budget. In particular, we find that when activities are substitutes, the planner has little incentive to intervene in more than two activities as long as all the spillovers are taken into account when deciding on the choice of intervention. This can have useful policy implications, as a planner does not have to spend resources in developing instruments for intervention in each activity. In this extension, we focus on the case where the interaction between each pair of activities is the same, and it would be interesting for future research to analyse  the \emph{key activity/layer problem} in  situations where the cross-activity interactions are heterogeneous.\footnote{We are grateful to Francis Bloch for this comment.}

We believe that the richer network structures offered by such multiplex networks can lead to further interesting results, and allow for the modelling of more economic scenarios. Our paper focuses on studying a formulation of the optimal intervention problem. Further research can be done to extend other results on single-activity networks to the multiple activity case, such as other issues in intervention, or to study network formation and design in multiplex networks. See, for instance, \cite{cheng2021theory}  on how to sustain cooperation among players embedded in multiple social relations, and  \cite{joshi2020network} and \cite{billand2021model}  on models of formation of multilayer networks.

\newpage
\bigskip
\appendix
\begin{center}{\Large \bf APPENDIX }\end{center}
\section{Two Useful Lemmas}\label{sect-A}
This section contains the proof of Lemma \ref{lem-1}, which characterizes the solutions to a general quadratic programming problem. For completeness, we include Lemma \ref{lem-3} as an extension of Lemma \ref{lem-1} to the case of intermediate budgets, which largely follows \cite{ggg}.

\textbf{Proof of Lemma \ref{lem-1}.}
Let \(\b y=C^{-1/2}\b x\), so we solve \begin{align*}
    \max_{\mathbf y\in\mathbb R^{2n}} \quad &C\b y^T\b S\b y+\sqrt C\hb v^T\b y\\
    \text{s.t.} \quad 
    &\b y^T\b y\leq 1.
\end{align*}
In general, we can solve for the optimal \(\b y\) by maximizing the Lagrangian \[\c L=C\b y^T\b S\b y+\sqrt C\hb v^T\b y+\lambda(\b y^T\b y-1).\]
This has first order conditions \begin{align*}2C\b S\b y+\sqrt C\hb v+2\lambda\b y&=0\\
\b y^T\b y&=1,\end{align*}
but there is no closed form solution to the system. We thus focus our analysis on the cases \(C\to\infty\) and \(C\to0\).\\
(a) We bound each term of the objective function individually. It is known that the problem 
\begin{align*}
    \max_{\mathbf y\in\mathbb R^{2n}} \quad &C\b y^T\b S\b y\\
    \text{s.t.} \quad 
    &\b y^T\b y\leq 1.
\end{align*}
has a maximum of \(\lmax(\b S)C\), which is attained for any \(\b y\in E_{max}(\b S)\) such that \(\|\b y\|=1\).\\
Also, for any \(\b y\) with \(\b y^T\b y\leq 1\), we have \(|\sqrt C\b v^T\b y|\leq\sqrt C\|\b v\|\|\b y\|\).\\ Therefore, \(V^*\in[\lmax(\b S)(C-\sqrt C\|\b v\|),\ \lmax(\b S)(C+\sqrt C\|\b v\|)].\)
Since \(\|\b v\|\) is a fixed constant, we have \(\lim_{C\to\infty}\frac{V^*}{C}=\lmax(\b S)\).\\
Also, \(V^*(\sqrt C\b u)\geq\lmax(\b S)(C-\sqrt C\|\b v\|)\) for all unit \(\b u\in E_{max}(\b S)\), so \(\lim_{C\to\infty}\frac{V^*(\sqrt C\b u)}{V^*(\b x^*)}=1\).\\

(b) Since \(\b S\) is symmetric, we can write \(\b S=\b U\b D\b U^T\), where \(\b D\) is a diagonal matrix consisting of the eigenvalues of \(\b S\), and  \(\b U=[\b u_1,\b u_2,\cdots,\b u_{2n}]\) is an orthogonal matrix consisting of the corresponding eigenvectors. If \(\b S\) has only one eigenvalue the result is trivial, so let \(\theta\) be the second largest eigenvalue of \(\b S\).\\

Since \(\b U\) forms a basis of \(\mathbb{R}^{2n}\), we can uniquely write \(\frac{\b x}{\|\b x\|}=\sum_i^{2n}k_i\b u_i\) for some \(k_i\in[-1,1]\) and \(\sum_i^{2n}k^2_i=1\). Then \begin{align*}
\frac{\b x^T\b S\b x}{\|\b x\|^2}&=\left(\sum_{i=1}^{2n}k_i\b u_i\right)^T\b S\left(\sum_{i=1}^{2n}k_i\b u_i\right)\\
&=\left(\sum_i^{2n}k_i\b u_i\right)^T\left(\sum_{i=1}^{2n}k_id_{ii}\b u_i\right)\\
&=\sum_{i=1}^{2n}k^2_id_{ii}\b u_i^T\b u_i+\sum_{i\neq j}k_ik_jd_{jj}\b u_i^T\b u_j\\
&=\sum_{i=1}^{2n}k^2_id_{ii}+0\\
&\leq\sum_{\{\b u_i\in E_{max}(\b S)\}}k_i^2\lmax(\b S)+\sum_{\{\b u_i\notin E_{max}(\b S)\}}k_i^2\theta\\
&=\lmax(\b S)-\left(1-\sum_{\{\b u_i\in E_{max}(\b S)\}}k_i^2\right)(\lmax(\b S)-\theta)
\end{align*}
Let \(\b A\) be the submatrix of \(\b U\) containing the columns of \(\b U\) which lie in \(E_{max}(\b S)\). Then \(\b A\) is a projection matrix for \(E_{max}(\b S)\), so \begin{align*}
    \|\mathrm{proj}_{E_{max}(\b S)}\b x\|^2&=\|\b A(\b A^T\b A)^{-1}\b A^T\b x\|^2
    =\b x^T\b A\b A^T\b A\b A^T\b x
    =\b x^T\b A\b A^T\b x
    =\|\b A^T\b x\|^2\\
    &=\|\b x\|^2\sum_{\b u_i\in E_{max}(\b S)}\left(\b u_i^T\sum_{j=1}^{2n}k_i\b u_j\right)^2\\
    &=\|\b x\|^2\sum_{\b u_i\in E_{max}(\b S)}k_i^2
\end{align*}
As \(C\to\infty\), from part (a), \(\frac{\b x^T\b S\b x}{\|\b x\|^2}\to\lmax(\b S)\), hence we have \[\sum_{\b u_i\in E_{max}(\b S)}k^2_i=\frac{\|\mathrm{proj}_{E_{max}(\b S)}\b x\|^2}{\|\b x\|^2}\to 1.\]\\

(c) We again bound each term of the objective function individually. Following part (a), we have \(C\b y^T\b S\b y\in[-\lmax(\b S) C,\lmax(\b S) C]\), while \(\sqrt C\b v^T\b y\) has maximum \(\sqrt C\|\b v\|\). Thus \(V^*\in[\sqrt C\|\b v\|-\lmax(\b S) C, \sqrt C\|\b v\|+\lmax(\b S) C].\) Given that \(\|\b v\|>0\), we have \(\lim_{C\to0}\frac{V(\b x^*)}{\sqrt C}=\|\b v\|\).\\

(d) Let \(\epsilon>0\). For any \(C<\frac{\epsilon\|\b v\|}{2\lmax(\b S)}\), we must have \begin{align*}&V(\b y^*)\geq\sqrt{C}\|\b v\|-\lmax(\b S) C\\
    \implies&C(\b y^*)^T\b S\b y^*+\sqrt C\b v^T\b y^*\geq\sqrt{C}\|\b v\|-\lmax(\b S) C\\
    \implies&\lmax(\b S) C +\sqrt C\b v^T\b y^*\geq\sqrt{C}\|\b v\|-\lmax(\b S) C\\
    \implies&\b v^T\b y\geq\|\b v\|-2\lmax(\b S)\sqrt C\\
    \implies&\frac{\|\text{proj}_{\b v}\b y\|}{\|\b y\|}=\frac{\b v^T\b y}{\|\b v\|\|\b y\|}\geq1-\frac{2\lmax(\b S)\sqrt C}{\|\b v\|}>1-\epsilon.
\end{align*}
Since \(\b y\) is a scalar multiple of \(\b x\), the desired result follows.

\begin{lemma}\label{lem-3}
Let \(\b S\) be a positive definite matrix. Let \(\b S=\b U\b D\b U^T\) be a diagonalization of \(\b S\), so \(d_{ii}=\lambda_i(\b S)\) are the eigenvalues of \(\b S\). Write \(\bb v=\b U^T\b v\), \(\bb x=\b U^T\b x\). The solution to \begin{align*}
    \max_{\mathbf x\in\mathbb R^{n}} \quad &V(\b x)=\mathbf x^T\mathbf{Sx}+\b v^T\b x\\
    \text{s.t.} \quad 
    &\b x^T\b x\leq C.
\end{align*}
is \[\bar x_i=\frac{\bar v_i}{2(\mu-\lambda_i(\b S))},\]
where \(\mu\) satisfies the equation \[\sum_{i=1}^n\frac{\bar v_i^2}{4(\mu-\lambda_i(\b S))^2}=C.\]
\end{lemma}

\textbf{Proof of Lemma \ref{lem-3}.} We have \(\b x^T\b S\b x=\bb x^T\b D\bb x=\sum_{i=1}^n\lambda_i(\b S)\bar x_i^2,\)
so the original problem can be reformulated as 
\begin{align*}
    \max_{\bb x\in\mathbb R^{n}} \quad & \sum_{i=1}^n\left(\lambda_i(\b S)\bar x_i^2+\bar v_i\bar x_i\right)\\
    \text{s.t.} \quad 
    &\sum_{i=1}^n \bar x_i^2\leq C.
\end{align*}
Consider the Lagrangian \[L=\sum_{i=1}^n\left(\lambda_i(\b S)\bar x_i^2+\bar v_i\bar x_i\right)+\mu(C-\sum_{i=1}^n \bar x_i^2).\]
The first order conditions are \[\frac{\partial L}{\partial\bar x_i}=2\lambda_i(\b S)\bar x_i+\bar v_i-2\mu\bar x_i=0\ \forall i,\]
so \[\bar x_i=\frac{\bar v_i}{2(\mu-\lambda_i(\b S))}.\]
The constraint \(\b x^T\b x\leq C\) thus implies that \(\mu\) satisfies the equation \[\sum_{i=1}^n\frac{\bar v_i^2}{4(\mu-\lambda_i(\b S))^2}=C.\]

When applied to the quadratic program \eqref{eq-objective}, Lemma \ref{lem-3} implicitly determines the optimal intervention \(\b x^*\) and welfare \(W(\b x^*)\) in terms of the spectral decomposition of \(\b P\).

\section{Proofs}\label{sect-B}
\textbf{Proof of Proposition \ref{pr-1}.}
Adapting from \cite{cyz}, the agents' choice \(\b x\) satisfies the first order conditions \[a_i^s-x_i^s-\sum_{t\neq s}\beta_{st}x_i^t+\delta\sum_{j\neq i}g_{ij}x_j^s=0\]
for all \(i\in\c N,\ s\in\c K\). In matrix form, this is \[\b 0=\b a-\b x-(\b \Phi\otimes \b I_n)\b x+(\b I_k\otimes\delta\b G)\b x=\b a-[\b I_k\otimes\b I_n+\b \Phi\otimes\b I_n-\b I_k\otimes\delta\b G]\b x,\]
so
\[\b x=[\tb \Phi\otimes\b I_n-\b I_k\otimes\delta\b G]^{-1}\b a.\]

Therefore, the total welfare across all agents is 
\begin{align*}
    W(\b a)&=\sum_{i\in\c N}\left(\sum_{s\in\c K} a_i^sx_i^s-\frac{1}{2}\sum_{s\in\c K}(x_i^s)^2-\frac{1}{2}\sum_{s\in\mathcal{K}}\sum_{\substack{t\in\mathcal{K}\\t\neq s}}\beta_{st}x_i^sx_i^t+\delta\sum_{i\in\mathcal{N}}\sum_{j\in\mathcal{N}}g_{ij}x_i^sx_j^s\right)\\
    &=\b a^T\b x-\frac{1}{2}\b x^T\b I_{kn}\b x-\frac{1}{2}\b x^T(\b \Phi\otimes\b I_n)\b x+\b x^T(\b I_k\otimes\delta\b G)\b x\\
    &=\b x^T\left[(\tb \Phi\otimes\b I_n-\b I_k\otimes\delta\b G)-\frac{1}{2}(\b I_k\otimes\b I_n)-\frac{1}{2}(\b \Phi\otimes\b I_n)+(\b I_k\otimes\delta\b G)\right]\b x\\
    &=\frac{1}{2}\b x^T(\tb \Phi\otimes\b I_n)\b x\\
    &=\frac{1}{2}\b a^T[\tb \Phi\otimes\b I_n-\b I_k\otimes\delta\b G]^{-1}(\tb \Phi\otimes\b I_n)[\tb \Phi\otimes\b I_n-\b I_k\otimes\delta\b G]^{-1}\b a.
\end{align*}

\textbf{Proof of Theorem \ref{th-1}.}
The expression in \eqref{eq-eigenvalues} is decreasing in \(\lambda(\tb\Phi)\) and increasing in \(\lambda(\delta\b G)\). Therefore, the expression is maximized at the smallest eigenvalue of \(\tb\Phi\) and the largest eigenvalues of \(\delta\b G\). Theorem \ref{th-1} then follows from an application of Lemma \ref{lem-1}.

\textbf{Proof of Proposition \ref{pr-2}.} Proposition \ref{pr-2} follows directly from Theorem \ref{th-1}(a).

\textbf{Proof of Corollary \ref{co-1}.} Since \(\lim_{C\to\infty}\rho(\b u\otimes\b v,\b a^*)=\pm 1\), we have \[\lim_{C\to\infty}\frac{B^s(\b a^*)}{C}=\lim_{C\to\infty}\frac{\sum_{i=1}^n(\b a^{s*}_i)^2}{C}=\lim_{C\to\infty}\frac{C\sum_{i=1}^n(u_sv_i)^2}{C}=(u_s)^2.\]
The other limits can be obtained similarly.

\textbf{Proof of Proposition \ref{pr-3}.} Let \(\b M=\b I_k-\b \Phi\). Then \(\b M\) is a nonnegative matrix, by the Perron-Frobenius theorem, the largest eigenvalue \(\lmax(\b M)\) has a corresponding nonnegative eigenvector \(\b u\). Note that \(\lmax(\b M)=1-\lmin(\b \Phi)\), so \(\b u\) is the desired nonnegative eigenvector in \(E_{min}(\b \Phi)\).

\textbf{Proof of Proposition \ref{pr-4}.} Here \(\b \Phi\) is a nonnegative matrix, so by the Perron-Frobenius theorem, the largest eigenvalue is at least the smallest row sum, which is 1. Furthermore, tr\((\b \Phi)=k\). Suppose \(\lmax(\b \Phi)=1\), then all the eigenvalues of \(\b \Phi\) are 1, but that means that \(\beta_{st}=0\) for all \(s,t\), a contradiction. Thus \(\lmax(\b \Phi)>1\), and \(\lmin(\b \Phi)<1\). Let \(\b u\) be an element of \(E_{min}(\b \Phi)\). If \(\b u\geq\b 0\), then \(\lmin(\b \Phi)\b u=\b \Phi\b u\geq\b I_k\b u=\b u\), a contradiction. Thus \(\b u\) must have both positive and negative components.

\textbf{Proof of Proposition \ref{pr-5}.} Since \(\beta_{st}\leq0\) for all \(s,t\), by the Perron-Frobenius theorem, we have that \(\lmax(-\b \Phi)=\lmin(\b \Phi)\) is weakly monotone in its entries, so \(\lmin(\b \Phi')\leq\lmin(\b \Phi)\). By Proposition \ref{pr-2}, \(R(\b \Phi,\delta\b G)\) is decreasing in \(\lmin(\b \Phi)\), so \(R(\b \Phi',\delta\b G)\geq R(\b \Phi,\delta\b G)\).

\textbf{Proof of Proposition \ref{pr-6}.} This result follows from an application of parts (c) and (d) of Lemma \ref{lem-1} on the optimization problem \eqref{eq-objective}.

\textbf{Proof of Proposition \ref{pr-7}.} Let \[\mathbf M_+=[(1+(k-1)\beta)\mathbf I_n-\delta \mathbf G]^{-1},\text{ and }\mathbf M_-=[(1-\beta)\mathbf I_n-\delta \mathbf G]^{-1}.\] Then (see \cite{cyz})\[\b P^{\mathcal{LL}}=\frac{1-\beta}{2}\b I_l\otimes\b M_-^2+\frac{1}{2k}\b J_l\otimes[(1+(k-1)\beta)\b M_+^2-(1-\beta)\b M_-^2].\]
We can apply a similar method as in Theorem \ref{th-1} (see \eqref{eq-eigenvalues}) to obtain the eigenvalues of \(\b P^{\mathcal{LL}}\) as \[\frac{\lambda_j(\b J_l)(1+(k-1)\beta)}{2k(1+(k-1)\beta-\lambda_i(\delta\b G))^2}+\frac{(k-\lambda_j(\b J_l))(1-\beta)}{2k(1-\beta-\lambda_i(\delta\b G))^2}\text{ for }i=1\cdots n,\ j=1\cdots l.\]

When \(l>1\), the eigenvalues of \(\b J_l\) are \(l\) and \(0\), so the eigenvalues of \(\b P^{\c {LL}}\) are \[\frac{l(1+(k-1)\beta)}{2k(1+(k-1)\beta-\lambda_i(\delta\b G))^2}+\frac{(k-l)(1-\beta)}{2k(1-\beta-\lambda_i(\delta\b G))^2}\text{ and }\frac{1-\beta}{2(1-\beta-\lambda_i(\delta\b G))^2}\] for \(1\leq i\leq n\).\\
When \(l=1\), the unique eigenvalue of \(\b J_l\) is 1, so the eigenvalues of \(\b P^{\c{LL}}\) are
\[\frac{1+(k-1)\beta}{2k(1+(k-1)\beta-\lambda_i(\delta\b G))^2}+\frac{(k-1)(1-\beta)}{2k(1-\beta-\lambda_i(\delta\b G))^2}\] for \(1\leq i\leq n\).

To obtain the largest eigenvalue in each case, define the function \(f(a,b)\coloneqq\frac{1+a}{2(1+a-b)^2}\) on the domain \(a>-1\), and \(a+1>b\). Then
\[\frac{\partial f}{\partial a}=-\frac{1+a+b}{2(1+a-b)^3}<0,\ \frac{\partial f}{\partial b}=\frac{(1+a)}{(1+a-b)^3}>0,\]
so \(f\) is decreasing in \(a\) and increasing in \(b\).\\
When \(l=1\), the eigenvalues of \(\b P^{\c{LL}}\) are \(f(-\beta,\lambda_i(\delta\b G))\), which is thus maximized when \(\lambda_i(\delta\b G)\) is maximum, so the largest eigenvalue is \(f(-\beta,\lmax(\delta\b G))\).\\
When \(l>1\), the eigenvalues of \(\b P^{\c{LL}}\) are \(f(-\beta,\lambda_i(\delta\b G))\) and \(\frac{l}{k}f((k-1)\beta,\lambda_i(\delta\b G))+\frac{k-l}{k}f(-\beta,\lambda_i(\delta\b G))\). The latter is larger if and only if \(f((k-1)\beta,\lambda_i(\delta\b G))>f(-\beta,\lambda_i(\delta\b G))\), which is equivalent to the condition \(\beta<0\) since \(f\) is decreasing in its first argument. Thus the largest eigenvalue is \(f(-\beta,\lmax(\delta\b G))\) when \(\beta>0\), and is \(\frac{l}{k}f((k-1)\beta,\lmax(\delta\b G))+\frac{k-l}{k}f(-\beta,\lmax(\delta\b G))\) when \(\beta<0\).

\textbf{Proof of Theorem \ref{th-2}.}
Suppose \(\beta<0\), and let \(\b u\in E_{max}(\delta \b G),\ \b v\in E_{max}(\b J_l)\). Then we have \begin{align*}\b P^{\c {LL}}(\b v\otimes\b u)&=\frac{1-\beta}{2}\b v\otimes\frac{1}{(1-\beta-\delta\lambda_j)^2}\b u+\frac{l}{2k}\b v\otimes\left(\frac{1+(k-1)\beta)}{(1+(k-1)\beta-\lambda_j))^2}-\frac{1-\beta}{(1-\beta-\lambda_j)^2}\right)\b u\\&=\lmax(\b P^{\c {LL}})(\b v\otimes\b u),\end{align*}
so \(\b v\otimes\b u\) is in the eigenspace of \(\lmax(\b P^{\c {LL}})\). By Lemma \ref{lem-1}, \(\b v\otimes\b u\) is asymptotically optimal. Furthermore, we know that \(E_{max}(\b J_l)=\text{span}\{\b 1_l\}\), which means that the planner should choose to conduct the same intervention \(t\b u\) across all activities for some scalar \(t\).

Suppose \(\beta>0\) and  \(l>1\) instead, so we choose \(\b u\in E_{max}(\delta \b G),\ \b v\in E_{min}(\b J_l)\). A similar expansion of \(\b P^{\c {LL}}(\b v\otimes\b u)\) implies that \(\b v\otimes\b u\) is in the eigenspace of \(\lmax(\b P^{\c {LL}})\). Since \(-1\) is an eigenvalue of \(\b J_l\) with multiplicity \(l-1\), the choice of \(\b v\) is no longer unique. In particular, \((1,-1,0,\cdots,0)^T\) is a possible choice of \(\b v\), giving the simple intervention we constructed.

{\bf Proof of Theorem \ref{th-3}.} 
Theorem \ref{th-3} follows directly from Proposition \ref{pr-7}.

\textbf{Proof of Theorem \ref{th-4}.}
Applying Lemma \ref{lem-1} to \eqref{eq-4}(c), we obtain that for all \(1\leq l\leq k\), \[\lim_{C\to0}\frac{W^*(l,C)-\widehat W}{\sqrt C}=2\left\|\begin{bmatrix}\b P^{\c{LL}}&\b P^{\c{LH}}\end{bmatrix}\hb a\right\|.\] This gives the first part of Theorem \ref{th-4}.\\

Now suppose \(\hb a^s=\hb a^t\) for all \(s,t\). Then for all \(l\), we have \(\b P^{\c L}\hb a=\b 1_l\otimes\b P^{\{1\}}\hb a\), so \(\|\b P^{\c L}\hb a\|=\sqrt l\|\b P^{\{1\}}\hb a\|\propto\sqrt l\), and hence \(\lim_{C\to0}\eta_k(l,C)=\sqrt{\frac{l}{k}}\).

\textbf{Proof of Proposition \ref{pr-8}.} Proposition \ref{pr-8} follows directly from Proposition \ref{pr-7}(a).

\textbf{Proof of Proposition \ref{pr-9}.}
From the proof of Proposition \ref{pr-7}, the function \(f(a,b)=\frac{1+a}{2(1+a-b)^2}\) is decreasing in \(a\). Then  \[\alpha=\frac{f(k-1)\beta,\lmax(\delta\b G)}{-\beta,\lmax(\delta\b G)}\] is decreasing in \(\beta\). Therefore, by Theorem \ref{th-3}, when \(\beta<0\), the welfare improvement ratio \(\lim_{C\to\infty}\eta_k(l,C)\) is increasing in \(\beta\), while when \(\beta>0\), the welfare improvement ratio \(\lim_{C\to\infty}\eta_k(1,C)\) is decreasing in \(\beta\) instead.

\textbf{Proof of Proposition \ref{pr-10}.}  Proposition \ref{pr-10} follows directly from Fact \ref{fact-1} and Proposition \ref{pr-9}(i).

\textbf{Proof of Proposition \ref{pr-12}.}
By Lemma \ref{lem-1}, the optimal intervention corresponds to \(\lambda_{max}(\b P)=\frac{1}{\lambda_{min}(\b P^{-1})}\) since Assumptions \ref{ass-1} and \ref{ass-4} imply that \(\b P\) is positive definite. By Proposition \ref{pr-11}, \(\lambda_{min}(\b P^{-1})\) is equal to \(\min_t \lambda_{min}[\b M(t)]\) as \(t\) varies across the eigenvalues of \(\b G\). Choose any \(\b x\in\mathbb R^{k}\) with \(\b x^T\b x=1\). Then \[\frac{d(\b x^T[\b M(t)]\b x)}{dt}=\b x^T(-2\b\Delta+2t\b\Delta\tb\Phi^{-1}\b\Delta)\b x.\]
Let \(\b S=\b\Delta-t\b\Delta\tb\Phi^{-1}\b\Delta\), so that \(\frac{d(\b x^T[\b M(t)]\b x)}{dt}=-2\b x^T\b S\b x\).

(a) Define the square root of \(\b\Delta\) as \(\b\Delta^{1/2}=\text{diag}(\sqrt{\delta^1},\cdots,\sqrt{\delta^k})\), so that we have \(\b S=\b\Delta^{1/2}(\b I-\lambda\b\Delta^{1/2}\tb\Phi^{-1}\b\Delta^{1/2})\b\Delta^{1/2}\).

When \(\lmin(\b G)\leq t\leq0\), it is clear that \[\b\Delta^{1/2}\tb\Phi^{-1}\b\Delta^{1/2}\succ0\implies \b I_k-t\b\Delta^{1/2}\tb\Phi^{-1}\b\Delta^{1/2}\succ0\\\implies\b S\succ 0.\]
Suppose instead \(0<t\leq\lmax(\b G)\). Then from Assumption \ref{ass-4}, \begin{align*}\tb\Phi-t\b\Delta\succ 0&\implies\b\Delta^{-1/2}\tb\Phi\b\Delta^{-1/2}\succ t\b I_k\implies\lmin(\b\Delta^{-1/2}\tb\Phi\b\Delta^{-1/2})>t\\&\implies\lmax(\b\Delta^{-1/2}\tb\Phi\b\Delta^{-1/2})<\frac{1}{t}\implies \b\Delta^{1/2}\tb\Phi^{-1}\b\Delta^{1/2}\prec\frac{1}{t}\b I_k\\&\implies\b I_k-t\b\Delta^{1/2}\tb\Phi^{-1}\b\Delta^{1/2}\succ0\implies\b S\succ 0.\end{align*}
Thus for all \(t\in[\lmin(\b G),\lmax(\b G)]\) and \(\b x\neq 0\), we have \(\b x^T\b S\b x>0\), and \(\b x^T[\b M(t)]\b x\) is decreasing in \(t\). Therefore, \(\min_{\b x} \b x^T[\b M(t)]\b x=\lambda_{min}(\b M(t))\) is decreasing in \(t\), so \(\lmin(\b P^{-1})\) is obtained when \(t=\lambda_{max}(\b G)\) and \(\mu=\lambda_{min}(\b M(t))\), with corresponding eigenvector \(\b u\otimes\b v\) satisfying \(\b u\in E_{min}(\b M(t))\) and \(\b v\in E_{max}(\b G)\).

(b) Define \(\b D=-\b\Delta\), then \(\frac{d(\b x^T[\b M(t)]\b x)}{dt}=2\b x^T(\b D+t\b D\tb\Phi^{-1}\b D)\b x\). A similar argument as in part (a) shows that \(\b D+t\b D\tb\Phi^{-1}\b D\) is positive definite, hence \(\b x^T[\b M(t)]\b x\) is increasing in \(t\) and the optimal intervention \(\b u\otimes\b v\) satisfies \(\b u\in E_{min}(\b M(t))\) and \(\b v\in E_{min}(\b G)\).

\textbf{Proof of Proposition \ref{pr-11}.}
We have \begin{align*}\c P^{-1}&=[\tb \Phi\otimes\b I_n-\b \Delta\otimes\b G](\tb \Phi^{-1}\otimes\b I_n)[\tb \Phi\otimes\b I_n-\b \Delta\otimes\b G]\\&=[\tb\Phi\otimes\b I_n-2\b\Delta\otimes\b G+\b\Delta\tb\Phi^{-1}\b\Delta\otimes\b G^2],\end{align*}
so for any \(\b u_{ji},\b v_i\), 
\begin{align*}\c P^{-1}(\b u_{ji}\otimes\b v_i)&=\tb\Phi\b u_{ji}\otimes\b I_n\b v_i-2\b\Delta\b u_{ji}\otimes\b G\b v_i+\b\Delta\tb\Phi^{-1}\b\Delta\b u_{ji}\otimes\b G^2\b v_i\\&=\tb\Phi\b u_{ji}\otimes\b v_i-2t\b\Delta\b u_{ji}\otimes\b v_i+t^2\b\Delta\tb\Phi^{-1}\b\Delta\b u_{ji}\otimes\b v_i\\&=\b M(t)\b u_{ji}\otimes\b v_i\\&=\mu\b u_{ji}\otimes\b v_i.\end{align*}
Hence \((\mu_{ji}^{-1},\b u_{ji}\otimes\b v_i)\) is an eigenpair of \(\c P\). It remains to show that these eigenvectors are orthogonal. Take any \(\b u_{ji}\otimes\b v_i\) and \(\b u_{lk}\otimes\b v_k\). If \(i=k\), then by construction, \(\b u_{ji}\) and \(\b u_{lk}\) are orthogonal eigenvectors of \(\b M(t_i)\), so \((\b u_{ji}\otimes\b v_i)^T(\b u_{lk}\otimes\b v_k)=0\times\b v_i^T\b v_i=0\). Otherwise \(i\neq k\), so \(\b v_i\) and \(\b v_k\) are orthogonal vectors of \(\b G\) and \((\b u_{ji}\otimes\b v_i)^T(\b u_{lk}\otimes\b v_k)=\b u_{ji}^T\b u_{lk}\times0=0\).

\textbf{Proof of Proposition \ref{pr-positive}.} Since \(\b a^*_i=0\) for all \(i\notin S\), we have \((\b a^*)^T\b P\b a^*=(\b a^*_S)^T\b P_S\b a^*_S\). Furthermore, since \(\b a^*_S\) is strictly positive, it must be a local maximum of the function \(f(\b z)=\b z^T\b P_S\b z\) in the domain \(\{\b z:\b z^T\b z\leq C\}\). Therefore, \(\b a_S^*\) must satisfy \(\b a_S^*\in E_{max}(\b P_S)\), and \(\frac{(\b a^*)^T\b P\b a^*}{C}=\frac{(\b a^*_S)^T\b P_S\b a^*_S}{C}=\lmax(\b P_S)\).

\textbf{Proof of Proposition \ref{pr-14}.} Let \(S=\{\b a:\b a\in\mathbb R^{kn},\ \b a^T\b a\leq C,\ \b a\geq\b 0.\}\). For any \(z\in\mathbb R\), we have \begin{align*}\max_{\b a\in S}\ \b a^T\b P\b a>z&\iff\exists(\b a\in S)\ \b a^T\b P\b a>z\\&\iff \exists(\b a\in S)\ \b a^T(\b P-\frac{z}{C}\b I)\b a>0\\&\iff\exists(\b a\in S)\ \b a^T(\frac{z}{C}\b I-\b P)\b a<0\\&\iff \frac{z}{C}\b I-\b P\notin CoPn, \end{align*}
where \(CoPn\) denotes the set of copositive matrices. By \cite{murty}, the problem of checking whether a symmetric matrix is copositive is an NP-hard problem. Therefore, the nonnegative intervention problem \(\max_{\b a\in S}\ \b a^T\b P\b a\) is also NP-hard.

\textbf{Proof of Proposition \ref{pr-13}.}
Under Assumption \ref{ass-4}, the matrix $\tb\Phi\otimes\b I_n-\b \Delta\otimes\b G$, which is symmetric, is positive definite, so does its inverse. Therefore, the profit $\pi(\b p)$, defined in \eqref{eq-firm}, is strictly concave in $\b p$. It suffices to check the first order conditions.
Differentiating the profit function in \eqref{eq-firm}, we obtain the first order condition \[\pi'(\b p^*)=[\tb\Phi\otimes\b I_n-\b \Delta\otimes\b G]^{-1}(\b a-\b p^*)-[\tb\Phi\otimes\b I_n-\b \Delta\otimes\b G]^{-1}(\b p^*-\b c)=0,\]
so \(\b a-\b p^*=\b p^*-\b c\), giving the optimal price vector \(\b p^*=\frac{\b a+\b c}{2}\), with corresponding profit following from substituting back into \eqref{eq-firm}. 

\newpage
\bibliographystyle{chicago}
\bibliography{bib-GGG-beta}

\end{document}